\documentclass{JHEP3}
\usepackage{graphicx,amsmath,amssymb}
\usepackage{epsfig,multicol}
\usepackage{epsf}
\usepackage{epstopdf}
\input{epsf.sty}

\usepackage{graphicx}
\usepackage{amssymb}
\usepackage{amsmath}

\def\p{\partial}

\def\half{{1\over 2}}
\def\({\left(}
\def\){\right)}
\def\[{\left[}
\def\]{\right]}

\def\e{\begin{equation}}
\def\q{\end{equation}}
\def\m{\begin{eqnarray}}
\def\n{\end{eqnarray}}

\def\half{\frac{1}{2}}

\title{Coleman-de Luccia Tunneling and the Gibbons-Hawking Temperature}
\author{S.-H. Henry Tye\footnote{sht5@cornell.edu}, Daniel Wohns\footnote{dfw9@cornell.edu}~ and Yang Zhang\footnote{yz98@cornell.edu} \\
{\em Laboratory for Elementary Particle Physics,
 Cornell University, Ithaca, NY 14853, USA}}

\abstract{We study Coleman-de Luccia tunneling in some detail. We show that, for a single scalar field potential with a true 
and a false vacuum, there are four types of tunneling, depending on the properties of the potential. A general tunneling process involves a combination of thermal (Gibbons-Hawking temperature) fluctuation part way up the barrier followed by quantum tunneling. The thin-wall approximation is a special limit of the case (of only quantum tunneling) where 
inside the nucleation bubble is the true vacuum while the outside reaches the false vacuum. Hawking-Moss tunneling is the (only thermal fluctuation) limit of the case where the inside of the bubble 
does not reach the true vacuum at the moment of its creation, and the outside is cut off by the de Sitter horizon before it reaches the false vacuum. We estimate the tunneling rate for this case and find that the corrections 
to the Hawking-Moss formula can be large. In all cases, we see that the bounce of the Euclidean action decreases rapidly as the vacuum energy density increases, signaling that the tunneling 
is not exponentially suppressed. This phenomenon may be interpreted as a finite temperature effect due to the Gibbons-Hawking temperature of the de Sitter space. 
As an application, we discuss the implication of this tunneling property to the cosmic landscape.
}

\begin{document}

\section{Introduction}

Tunneling in the presence of gravity was first beautifully analyzed by Coleman and de Luccia (CDL) \cite{Coleman:1980aw}. This problem emerges in the study of the early universe in relation to the inflationary epoch, and more recently, in relation to the cosmic landscape as suggested by superstring theory. However, there are properties of the CDL tunneling that remain to be better understood.

\begin{figure}[ht]
\includegraphics{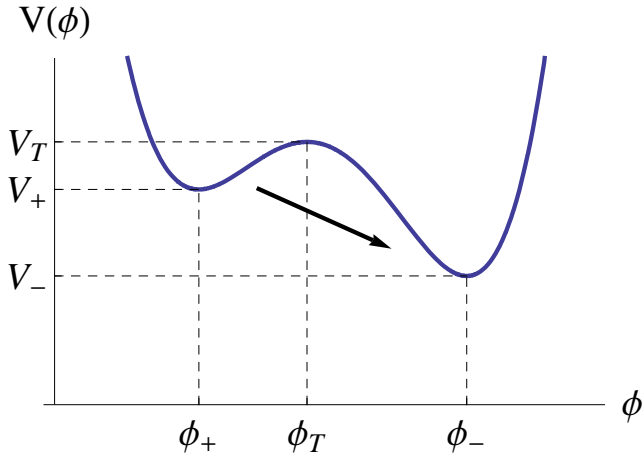}
\caption{Potential $V(\phi)$.
}
\label{potent}
\end{figure}

Consider a simplified theory of a single scalar field with potential $V(\phi)$ and a canonical kinetic term in the presence of gravity. The potential $V(\phi)$ has a false vacuum $V_+$ at $\phi_+$ and a true vacuum $V_-$ at $\phi_-$, so the energy density difference between them is $\epsilon= V_+-V_- >0$.
The potential barrier between these two local minima has height  $V_T$, as shown in Figure \ref{potent}.

In the semiclassical approximation, the tunneling rate  per unit volume is given by
\begin{equation}
\label{GammaB}
\Gamma \simeq A e^{-B},
\q 
\begin{equation}
B=S_E(\phi)-S_E(\phi_+) 
\label{bounce1}
\q 
where $S_E(\phi)$ is the Euclidean action for the bounce solution and $S_E(\phi_+)$ is the Euclidean action
evaluated at the false vacuum at $\phi_+$.  
In the thin-wall approximation ($\epsilon \rightarrow 0$), when the vacuum energy density is negligible compared to the Planck scale, $B$ recovers the result in the absence of gravity \cite{Coleman:1977py}, as expected. On the other hand,
as we increase $V_+ \gtrsim V_-$, while still in the thin-wall approximation \cite{Huang:}, $B$ takes a different form, 
\begin{equation}
B \simeq \frac{27\pi^2 \tau^4}{2 \epsilon^3} \rightarrow \frac{2\pi^2 \tau}{H^3}
\label{huangtw}
\q 
where $\tau$ is the domain wall tension and 
the Hubble constant is given by $H^2 = 8 \pi G_N V_+/3=V_+/3M_p^2$.
Note that the first form (when $V_+$ is small) is independent of $V_+$ while the second form is independent of $\epsilon=V_+ - V_-$. We see that, for fixed $\tau$, $B$ decreases rapidly as $H$ increases. It seems that we can easily have a situation where $B \ll 1$, in which case, the tunneling is not exponentially suppressed at all (and the evaluation of the prefactor $A$ in (\ref{GammaB}) as well as sub-dominant contributions to $\Gamma$ becomes important).

A similar phenomenon seems to be happening in the Hawking-Moss (HM) tunneling \cite{Hawking:1981fz},
\begin{equation}
B_{HM}= 24 \pi ^2 M_p^4 \bigg( \frac{1}{V_+} - \frac{1}{V_T} \bigg) =  24 \pi ^2 M_p^4\frac{\Delta V_+}{V_TV_+} \sim \frac{8 \pi \Delta V_+}{3 H^4} 
\label{BHM0}
\q
where $\Delta V_+= V_T - V_+$.
Note that if we move $V(\phi)$ up without changing its shape, that is, keeping $\epsilon$, $\tau$ and $\Delta V_+$ fixed,
$B$ decreases like $H^{-3}$ or $H^{-4}$, depending on which formula  (i.e., thin wall or HM) is applicable.
In this paper, we like to see if and under what conditions this phenomenon is real. More precisely, we examine the CDL tunneling more generally to see when the thin-wall approximation or the HM formula (\ref{BHM0}) is valid.

CDL tunneling concerns quantum fluctuations in the false vacuum of a nucleation bubble which subsequently grows classically. The following picture emerges. 
Depending on the properties of $V(\phi)$, there are four regions in the parameter space, yielding four different situations how the nucleation bubble is created :

(I) The center of the nucleation bubble is in the true vacuum $V_-$ while the outside of the bubble reaches 
the false vacuum $V_+$. The thin-wall approximation is a special limit ($\epsilon \rightarrow 0$) in this case. 

(II) The outside of the bubble reaches the false vacuum $V_+$ but  the inside of the nucleation bubble never
reaches the true vacuum $V_-$ in the Euclidean solution. After the creation of the bubble, its inside falls towards the true vacuum $V_-$ as the bubble grows.

(III) The inside of the bubble reaches $V_-$, but the outside of the bubble does not reach the false vacuum in the Euclidean action, due to the presence of the de Sitter horizon. 

(IV) Not only does the inside of the bubble not reach $V_-$, the outside of the bubble does not reach the false vacuum in the Euclidean action, due to the presence of the de
Sitter horizon. HM tunneling is a special limit in this case. 

In the absence of gravity, regions faraway from the bubble are by definition in the false vacuum, so we have only case (I) and (II). In the presence of gravity, the de Sitter
horizon cut off the scalar field $\phi$ as it tries to reach the false vacuum at large distance from the nucleation bubble \cite{Jensen:Steinhardt}. That is why case (III) and (IV) are possible. 
This phenomenon of cases (III) and (IV) may be interpreted as a Gibbons-Hawking (GH) temperature $T_H =H/2\pi$ effect. This point was recently made by Brown and Weinberg \cite{Brown},
who show that tunneling in the presence of gravity may be interpreted as a combination of thermal (Gibbons-Hawking temperature) fluctuation plus quantum tunneling. We agree with this picture. That is, $\phi$ thermally fluctuates from the false vacuum at $\phi_+$ part way up the potential (to $\phi_{f+}$) before quantum tunneling to the other side of the barrier. In the Hawking-Moss limit, it fluctuates all the way to the top of the barrier at $V_T$. Here, our explicit calculations (in particular, the determination of  $\phi_{f+}$) allow us to estimate in some detail, for a given potential,  the individual contribution of the thermal fluctuation versus that of the quantum tunneling in the tunneling rate. We also evaluate the back-reaction effect on the background geometry and show that it can be very important.

There are a few other properties that are worth mentioning here.

$\bullet$ The thin-wall approximation is valid only when the tunneling belongs to case (I). In addition to the case (I) conditions (to be specified), we see that $M_P^4 \gg V_T\gg \Delta V_+ \gg \epsilon$. As we increase $V_T$ with a generic fixed shape of $V(\phi)$ (with $\epsilon >0$), the tunneling goes over to case (IV) via case (II) and/or case (III).

$\bullet$ For large $V_T$ and $\epsilon >0$, case (IV) is generic, where HM tunneling is a special limit. Here we estimate the accuracy of the HM formula (\ref{BHM0}) and its correction. We see that the HM formula for $B$ is an over-estimate, by as much as a factor of three. That is, the actual decay is a combination of thermal tunneling and quantum tunneling, whose rate can be much faster.

$\bullet$ Another interesting point is that tunneling in the absence of gravity is always downwards, while gravitational effects allow tunneling upwards.

$\bullet$ With an overall picture, we see that the fast drop off of $B$ as $H$ increases is real. This is illustrated in 
Figure \ref{power}, where case (III) does not appear due to the particular choice of the potential used.

$\bullet$ Strictly speaking, for a ``general" smooth potential with varying
parameters, the above four cases may reduce to case (IV) only. However, within case (IV), it contains regions that resemble the various cases discussed above. The triangular potential has four different cases so it actually demonstrates the change from thin wall to HM case more dramatically (but does not change the physics qualitatively.)

$\bullet$ As $B$ decreases, multi-nucleation bubbles will form and the phase transition involves bubble collisions.
Once $B$ is small (say $B \lesssim 1$), Eq.(\ref{GammaB}) is no longer valid. The prefactor and sub-leading terms will become important. All we can say is that tunneling is no longer exponentially suppressed. The rate of transition depends on the details. We do expect that O(4) nucleation bubbles no longer dominate. That is, bubbles of other shapes as well as bubble collisions become important.  The transition can be complicated and is at a fast time scale. Something analogous to spinodal decomposition may happen. 

\begin{figure}
\begin{center}
\includegraphics[width=10cm]{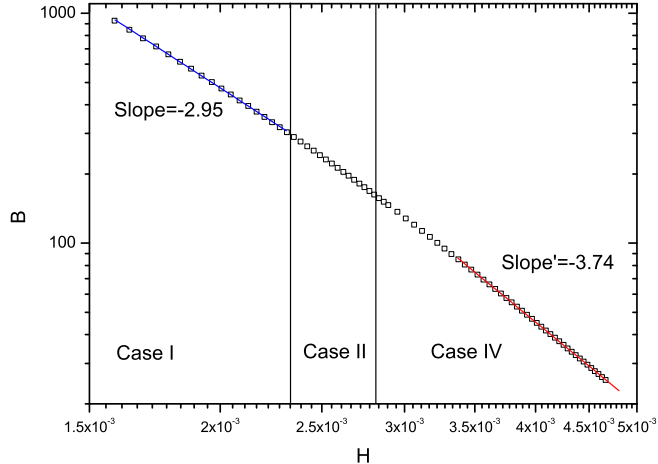}
\end{center}
\caption{The log-log plot of the tunneling exponent $B$ as a function of the Hubble parameter $H$ for a fixed potential $V(\phi)$ except for its overall height as measured by $H$. ($G\equiv 1$). The three cases are separated by the two solid vertical lines. As $H \sim \sqrt{V_T}$ increases, the tunneling behavior goes from case (I) to case (IV) via case (II); that is, it goes from the CDL thin-wall approximation (to the left in case (I), with slope $\sim -3$ as given by Eq.(\protect\ref{huangtw})), to the HM tunneling (to the right of case (IV), with slope $\sim -4$ as given by Eq.(\protect\ref{BHM0})). For this particular potential, the slope case (IV) ($\sim -3.74$) never quite reaches the HM value due to corrections. The specific potential used here is described in Sec. 6. Case (III) does not appear for this potential.}
\label{power}
\end{figure}

In terms of pure thermal (Gibbons-Hawking (GH) temperature) tunneling
the suppression of the tunneling is interpreted as due to the Boltzmann factor, where the bubble is a 3-sphere and the inverse temperature is treated as an imaginary time with
period $1/T_H$. In case (I) and (II), we see that the quantum tunneling is dominant (i.e., has a smaller value) due to the $O(4)$ symmetry that is lacking in the finite temperature effect. So strictly speaking, 
the enhancement of the tunneling rate due to a large vacuum energy density is a pure gravitational effect. In de Sitter-like vacua, we may interpret this as a GH temperature effect in the presence of an enhanced (here $O(4)$) symmetry. 
In case (III) and (IV), the GH temperature starts to play a more prominent role.

The same GH temperature will also contribute to the finite temperature effect on the potential $V(\phi) \rightarrow V(\phi, T_H)$ \cite{Allen,Kane}. This contribution is perturbative in the couplings while its effect in tunneling is non-perturbative. In the finite temperature formalism, this leads to a term of the form $T_H^2 \phi^2$ into the finite temperature potential. In the gravity perspective, this is simply a coupling of the form $R \phi^2$, where $R$ is the Ricci scalar. One an also interpret this as a finite volume effect, due to Gauss's law, $H^2 \phi^2$, where $H^{-1}$ is the horizon size. The overall picture is self-consistent and clear.

We shall comment on the impact of the GH temperature on the cosmic landscape, both on the shape of the effective potential of the landscape and on the tunneling rate. Because of fast tunneling when the wavefunction of the universe is high up in the landscape, it is likely that the universe is quite mobile there.

For a general potential $V(\phi)$, the coupled equations of $\phi(\xi)$ and the cosmic scale factor $r(\xi)$ are too complicated to solve even in the absence of the gravity 
except in the thin-wall approximation.  
However, for some special potentials, say, the triangular potential $V(\phi)$ (Figure \ref{Triangle}), the bounce $B$ can be obtained analytically \cite{Duncan:Jensen}. So it is
natural to consider the tunneling rates for such a potential in the presence of gravity. Unlike Ref.\cite{Duncan:Jensen}, we cannot get an exact analytic formula for $B$ in the presence of gravity, but the resulting analytic study gives a very good approximation and does simplify enough for us to see the overall picture. Here
the absolute height of the potential, which corresponds to the vacuum energy density, is important for tunneling with gravity. To be specific, we shall adopt this triangular potential in this paper.

The rest of the paper is organized as follows. In Sec. 2, we briefly review the overall framework of tunneling in de Sitter space. This framework is the CDL tunneling formalism. In Sec. 3,
we go back to tunneling in the absence of gravity. In particular, we review the special case of tunneling in a triangular potential studied by Duncan and Jensen
\cite{Duncan:Jensen}. Here we emphasize the formulation that is suitable in the extension of their analysis to include gravity. Sec. 4  presents the setup for tunneling in de
Sitter space, again using the triangular potential. As $\phi$ varies, the tunneling is not happening in pure de Sitter space and the back-reacrtion is estimated. Sec. 5 presents
the main result of this paper. The above four cases and their conditions are discussed. The meaning and implications of the results are discussed in Sec. 6. We then discuss
thermal tunneling in Sec. 7. Here we are referring to the treatment of the de Sitter horizon effect as a Gibbons-Hawking temperature effect. In Sec. 8, we point out that the
Gibbons-Hawking temperature should also modify the potential via finite temperature effect on potentials. We then consider the implication of the Gibbons-Hawking temperature on the cosmic landscape. Sec. 9 contains the summary and some remarks. 
 


\section{Coleman-de Luccia Tunneling}

Let us consider the theory of a single scalar field $\phi$ with a potential $V(\phi)$ in the presence of gravity, given by
\begin{equation}
S=\int d^4x\sqrt{-g}\[\half
g^{\mu\nu}\p_\mu\phi\p_\nu\phi-V(\phi)-{R\over 2\kappa}\],
\q 
where $\kappa=8\pi G =1/M_p^2$.

The potential $V(\phi)$ has a false vacuum $V_+$ at $\phi_+$ and a true vacuum $V_-$ at $\phi_-$. 
There is a potential barrier between these two local minima, as shown in Figure \ref{potent}. Let the height of the barrier at 
$\phi_T$ be $V_T$. We have chosen $\phi_- > \phi_T >\phi_+$. Let the energy density difference between the false and the true vacua be $\epsilon= V_+-V_- >0$.
 
In the presence of gravity, the CDL tunneling rate per unit volume is given by Eq. (\ref{GammaB},\ref{bounce1}), 
in term of  the coefficient $B=S_E(\phi)-S_E(\phi_+)$, 
where $S_E(\phi)$ is the Euclidean action for the bounce solution and $S_E(\phi_+)$ is the Euclidean action
evaluated at the false vacuum. Since a Euclidean solution with a $O(4)$ symmetry has an etremum action, $B$ is
in general dominated by the ``bounce'' solution with $O(4)$ symmetry. This solution has the Euclidean metric,
\begin{equation}
ds^2=d\xi^2 +r(\xi)^2 d\Omega^2_s, 
\q   
where $d\Omega^2_s$ is the metric of a unit 3-sphere. The Euclidean equation for the ``bounce'' solution is 
determined by the minimum value of the Euclidean action, 
\m
\phi''+\frac{3 r'}{r} \phi'=\frac{dV}{d\phi} \label{eqphi} \\
r'^2=1+\frac{ r^2}{3 M_p^2}  \bigg(\frac{1}{2} \phi'^2-V \bigg).
\label{eqr}
\n
where the prime is derivative with respect to $\xi$.
Here the Einstein equation yields one non-trivial equation, since the equation of motion for $r(\xi)$ follows from Eqs. (\ref{eqphi},\ref{eqr}).
We can choose $r(0)=0$.
Using Eq.(\ref{eqr}) to simplify the Euclidean action, one obtains
\begin{equation}
 S_E=4\pi^2\int d\xi \[r^3 V -{3r \over \kappa}\]  \q

The CDL instanton is a unique solution with the topology of a four sphere $S^4$ in Euclidean space \cite{Mottola:Lapedes}.
The geometry after bubble
nucleation is described by the analytic continuation of the CDL
instanton to Lorentzian signature. The radial coordinate
$\xi$ is continued to $\xi=it$ and the metric in Lorentzian frame
is \begin{equation} -ds^2=-dt^2+r^2(it)d\Omega_{H^3}^2,\q here the metric is
multiplied by an overall minus sign and $d\Omega_{H^3}$ is the
element of length for a unit hyperboloid with timelike normal. The
metric within the spherical bubble describes a spatially open
Friedmann-Robertson-Walker Universe.

As we shall see, a qualitative picture emerges: depending on the properties of the potential,
there are four different cases how the nucleation bubble is created :

(I) The center of the nucleation bubble reaches the true vacuum $V_-$ while the outside of the bubble reaches 
the false vacuum $V_+$. The thin-wall approximation is a special limit ($\epsilon \rightarrow 0$) in this case. 

(II) The outside of the bubble reaches the false vacuum $V_+$ but  the inside of the nucleation bubble never
reaches the true vacuum $V_-$ in the Euclidean solution. After the creation of the bubble, its inside will fall towards the true vacuum $V_-$ as the bubble grows.

(III) The inside of the bubble reaches $V_-$, but the outside of the bubble does not reach the false vacuum $V_+$ in the Euclidean action, due to the presence of the de Sitter
horizon. This case and the next case never happen in the absence of gravity, since we start with the false vacuum everywhere. Far from the bubble, $\phi$ approaches the false
vacuum by definition. However, when the vacuum energy is not negligible, there is a de Sitter horizon so it is possible that $\phi$ hits the horizon before it reaches the false vacuum. 

(IV) Not only does the inside of the bubble not reach $V_-$, the outside of the bubble does not reach the false vacuum in the Euclidean action. The HM tunneling is the limit in this case. Here we are able to estimate the accuracy of the Hawking-Moss formula and its correction, which can be substantial.   

For a general potential $V(\phi)$, the coupled equations (\ref{eqphi}, \ref{eqr}) are complicated to solve.
In the absence of the gravity, for some simple potentials, say, the triangular potential $V(\phi)$ (Figure \ref{Triangle}), the bounce $B$ can be obtained analytically 
\cite{Duncan:Jensen}. Then it is natural to consider the tunneling rates for such potentials in the presence of the gravity. Unlike Ref.\cite{Duncan:Jensen}, we cannot get 
an analytical formula for $B$ in the presence of gravity, but the analysis does simplify enough for us to see the overall picture. Here
the absolute height of the potential, which corresponds to the vacuum energy density, is important for tunneling with gravity. To be specific, we shall adopt this triangular potential in this paper.

The triangular potential may be parametrized in the following way. Let the height of the barrier of $V(\phi)$ at $\phi_T$ be $V_T$, which also provides a measure of the overall vacuum energy density.  Let
$\Delta V_{\pm} = (V_T-  V_{\pm})$. Both $\Delta \phi_+=\phi_T-\phi_+$ and $\Delta \phi_-=\phi_--\phi_T$ are defined to be positive so the slopes (gradients)  $\pm \lambda_{\pm}$ of $V(\phi)$ are given by
\begin{equation}
\lambda_{\pm} = \frac{\Delta V_{\pm}}{\Delta \phi_{\pm}}
\label{lambda}
\q
Note that, in general, $\lambda_+ \ne \lambda_-$.
Since the $\epsilon =  {\Delta V_-} - {\Delta V_+} >0$, we have
\begin{equation}
\eta= \sqrt{\frac{\Delta V_+}{\Delta V_-}} = \frac{1}{\sqrt{1 + \epsilon/\Delta V_-}} <1
\label{eta}
\q

$\Delta \phi_+$, $\Delta \phi_-$ and $\lambda$. Here, to simplify the solution, we set $\lambda_L=\lambda_R=\lambda$. 
\begin{figure}[t]
\centering
\includegraphics{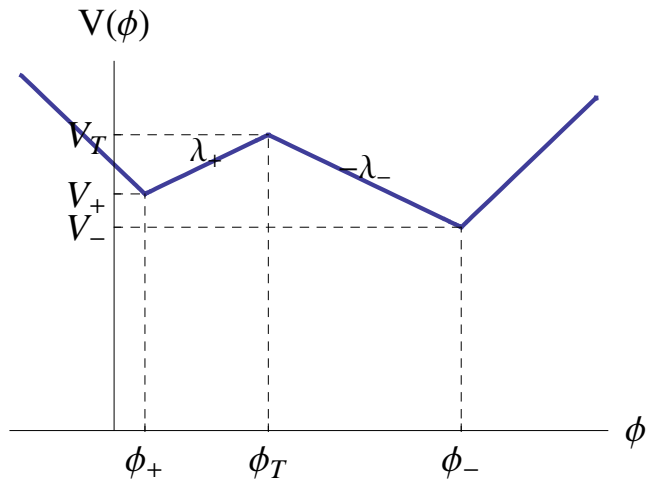}
\caption{Triangular potential $V(\phi)$. The false vacuum is at $V_+=V(\phi_+)$ and the true vacuum is at $V_-=V(\phi_-)$, with the top of the barrier at $V_T=V(\phi_T)$. Here $\lambda_+$ and $-\lambda_-$ are the gradients.}
\label{Triangle}
\end{figure}

\section{Tunneling Without Gravity}

When gravity is negligible, Eq. (\ref{eqr}) reduces to $r'=1$. This happens if we set $\kappa=0$. Alternatively, this is a very good approximation when the overall height of the potential is much smaller than the Planck scale. 

With $r(\xi=0)=0$, we have $r=\xi$, so tunneling reduces to the simple case without gravity. The solution must satisfy the boundary condition,
\begin{equation}
\phi'(0) =0
\label{cond1}
\q 
which is necessary for $\phi$ to make sense at the center of the nucleation bubble. 
At large radius, we expect
\begin{equation}
\phi \rightarrow \phi_+ ,       \quad \quad \xi \rightarrow \infty
\label{cond2}
\q
As we shall see, this condition may be modified when gravity is important.

Let us first review the solution to the triangular potential without gravity, the case worked out by Duncan and Jenson \cite{Duncan:Jensen}. For $\phi_+ \le \phi \le \phi_-$, Eq.(\ref{eqphi}) becomes
\begin{equation}
\phi''+\frac{3 }{\xi} \phi'=\frac{dV}{d\phi} = \pm \lambda_{\pm} 
\label{eqphitri} 
\q
The general solution is 
\begin{equation}
\phi(\xi) = a  + b/\xi^2 \pm \lambda_{\pm} \xi^2/8 
\q
where the constant $a$ and $b$ are determined by the boundary conditions and by matching the field values and their derivatives at the top of the barrier, which occurs at some radius $\xi_T$ to be determined.

It is easy to see that $\phi$ will reach its false vacuum value at finite radius $\xi_+$ (to be determined) and then stay there. The above boundary condition (\ref{cond2}) is replaced by
\begin{equation}
\left\{ \begin{array}{ll}
\phi(\xi_+)=\phi_+ \\
\phi'(\xi_+)=0
\end{array} \right.
\label{cond3}
\q
while there are two possibilities to satisfy the condition (\ref{cond1}). The first case, namely case (I), is when $\phi$ stays close to $\phi_-$ until at radius $\xi_-$ when $\phi$
starts to decrease under Eq.(\ref{eqphitri}). The boundary conditions in this case are
\begin{equation}
\left\{ \begin{array}{ll}
\phi(\xi)=\phi_-     \quad \quad 0 \le \xi \le \xi_- \\
\phi'(\xi_-)=0
\end{array} \right.
\label{cond4}
\q
The other possibility, namely case (II), happens when, at the time of creation, the inside of the bubble never reaches the true vacuum. In this case, the boundary conditions 
are  
\begin{equation}
\left\{ \begin{array}{ll}
\phi(0)=\phi_0 \\
\phi '(0)=0
\end{array} \right.
\end{equation}
where the initial value $\phi_0$ is to be determined. This case only happens if $\phi_0$ is to the right side of the barrier (i.e., near $\phi_-$) and $\phi_0 \le \phi_-$. Also, we expect $V_+ \ge V(\phi_0) \ge V_- $.

\subsection{Case (II)}

Let us consider case (II) first. On the right and left sides of the barrier, we have
\begin{equation}
\left\{ \begin{array}{ll}
\phi_R(\xi)=\phi_0   - \frac{\lambda_-}{8} \xi^2     \quad \quad  0 \le \xi \le \xi_T \\
\phi_L(\xi)= \phi_+ + \frac{\lambda_+}{8 \xi^2} (\xi^2-\xi_+^2)^2  \quad \quad  \xi_T \le \xi \le \xi_+
\end{array} \right.
\label{sol2}
\end{equation}
Matching the derivatives of the two solutions (\ref{sol2}) at $\xi_T$, 
\begin{equation}
\xi_+^4 = (1+c) \xi_T^4
\q
where $c=\lambda_-/\lambda_+$. Matching the field values at $\xi_T$ yields
\begin{equation}
\left\{ \begin{array}{ll}
\phi_0=\phi_T +   \frac{\lambda_-}{8} \xi_T^2   \\
\Delta \phi_+ = \phi_T - \phi_+=  \frac{\lambda_+}{8}\left(\sqrt{1 +c} -1\right)^2 \xi_T^2
\end{array} \right.
\label{solpar2}
\q
Now that the unknowns $\xi_T$, $\xi_+$ and $\phi_0$ are determined in terms of properties of the potential $V(\phi)$, we simply insert the solution (\ref{sol2}) into the Euclidean action and integrate it from $\xi=0$ to $\xi=\xi_+$.
The bounce (\ref{bounce1}) is given by
\begin{equation}
\label{b2}
B = \frac{32\pi^2}{3} \frac{1+c}{(\sqrt{1+c} -1)^4} \frac{(\Delta\phi_+)^4}{\Delta V_+}
\q
Let $\Delta V_0= V_T - V_0= V_T - V(\phi_0)$ so, using Eq.(\ref{solpar2}),
\begin{equation}
\frac{\Delta V_+}{\Delta V_-} < \frac{\Delta V_+}{\Delta V_0} = \frac{\left(\sqrt{1 +c} -1\right)^2}{c^2} 
\label{V0cond}
\q
which implies that tunneling is always downward, and  ${\Delta V_+}/{\Delta V_0} \le 1/4$.

Recall that case (II) holds only if $V_0>V_-$, or $\phi_0 \le \phi_-$, which translates to the above condition (\ref{V0cond}).
or equivalently, $\beta >1$, where
\begin{equation}
\beta = \sqrt{\frac{\Delta V_-}{\Delta V_+}} \frac{\Delta \phi_- -  \Delta \phi_+}{2\Delta \phi_-} 
\label{beta}
\q
In general, $\beta$ is semi-positive. 
If $\beta =1$, then $\phi_0=\phi_-$ and one can rewrite the bounce (\ref{b2}) in the following form
\begin{equation}
B= \frac{2 \pi^2}{3} \frac{(\Delta \phi_-^2-  \Delta \phi_+^2)^2}{\Delta V_+}
\label{b21}
\q 

\subsection{Case (I)}

If $\beta < 1$, the inside of the bubble reaches the true vacuum $V_-$, that is, $\phi(0)=\phi_-$. 
At certain radius $\xi_-$, 
$\phi$ begins to decrease until it reaches $\phi_+$ at radius $\xi_+$. The region for $\xi > \xi_+$ stays at the false vacuum.   Now, the solution 
$\phi(\xi)$ contains four pieces, 
\begin{equation}
\phi(\xi) = \left\{ \begin{array}{ll}
\phi_- & \textrm{$x\in [0,\xi_-]$} \\
\phi_1(\xi) =  \phi_- - \frac{\lambda_-}{8 \xi^2} (\xi^2-\xi_-^2)^2 & \textrm{$x\in [\xi_-,\xi_T]$}\\
\phi_2(\xi) = \phi_+ +  \frac{\lambda_+}{8 \xi^2} ( \xi^2-\xi_+^2)^2 & \textrm{$x\in [\xi_T,\xi_+]$}\\
\phi_+ &  \textrm{$x \ge \xi_+$}
\end{array} \right.
\label{sol1}
\end{equation}
where $\phi_- \ge \phi_1(\xi) \ge \phi_T$ and $\phi_+ \le \phi_2(\xi)\leq \phi_T$.
Now we have three unknowns : $\xi_T$, $\xi_-$ and $\xi_+$. Matching the derivatives as well as the field values at $\xi_T$ yields
\begin{equation}
\left\{ \begin{array}{ll}
\xi_+^4-\xi_T^4 = c(\xi_T^4-\xi_-^4) \\
\Delta \phi_+ =\frac{\lambda_+}{8 \xi_T^2} \left(\xi_T^2 - \xi_-^2 \right)^2 \\
\Delta \phi_- =\frac{\lambda_-}{8 \xi_T^2} \left(\xi_T^2 - \xi_-^2 \right)^2 
\end{array} \right.
\q
so $\xi_T$, $\xi_-$ and $\xi_+$ can be solved in terms of the properties of $V(\phi)$.
Once again, one can insert these solutions of the parameters into the $\phi$ solution (\ref{sol1}) and evaluate the bounce. We find that it is convenient to introduce the tension to be
\begin{equation}
\tau = \tau_+ + \tau_-
\q
\begin{equation}
\label{eq2}
\tau_{\pm} = \pm \int^{\phi_T}_{\phi_{\pm} } d\phi \sqrt{2[V(\phi) - V_{\pm}]} = \frac{2}{3} \sqrt{2 \Delta V_{\pm}} \Delta \phi_{\pm}
\q   
After some algebra,  $B$ can be written in terms of these values
\begin{equation}
\label{b1}
B = \frac{27\pi^2}{32 \epsilon^3} [ \tau_+(1+1/\eta) + \tau_-(1 + \eta)]^3 [\tau_+(3/\eta -1) +\tau_- ( 3\eta -1)]
\q
where $\epsilon=V_+ - V_-$ and $\eta^2=\Delta V_+/\Delta V_- <1$ (\ref{eta}). 
In the limit $\beta \rightarrow 1$, $B$ in (\ref{b1}) reduces to B in (\ref{b21}), that is, it agrees with B in case (II).

Here $\tau_{\pm}$ in Eq. (\ref{eq2}) should be treated as a convenient definition. The value of $\tau$ as defined should be very close to the actual domain wall tension.
For more general potentials, the coefficient $2/3$ in Eq. (\ref{eq2}) will have order one variations.
Note that $\epsilon/\Delta V_- = 1-\eta^2$.
In the thin-wall approximation, $\eta \rightarrow 1$ as $\epsilon/\Delta V_-$ becomes small, and $B$ reduces to
the usual thin-wall formula for $B$,
\begin{equation}
\label{b1tw}
B \rightarrow \frac{27\pi^2}{2 \epsilon^3} \tau^4
\q
So we see that the thin-wall approximation is inside the region of case (I). Comparing the small $\epsilon/\Delta V_-$ 
case to that in case (I) Eq.(\ref{V0cond}), we see that there is a sizable parameter region where the thin-wall approximation is not valid.
For case (II), since the inside of the bubble never reaches the true vacuum value $\phi_-$, the wall tension does not carry much physical significance here. The non thin-wall approximation has also been studied in Ref.\cite{Gen:1999gi}.

In summary, the regions (I) and (II) are divided by the value of $\beta$ (\ref{beta}):  case (I) if $0 \le \beta <1$ and case (II) 
($\phi(0) = \phi_0 < \phi_-$) if $\beta >1$. For $\beta=1$, only the center of the bubble reaches the true vacuum. 

In the absence of gravity, the overall height of the potential is not important, so the triangular potential is parameterized by a set of four parameters, namely 
($\Delta V_{\pm}$, $\Delta \lambda_{\pm}$). This set can be replaced by an equivalent set ($\Delta V_{\pm}$, $\Delta \phi_+$, $c$). The bounce $B$ in Eq.(\ref{b2}) is expressed 
in terms of this set, where we see that $\Delta V_-$ does not enter, since the inside of the bubble never reaches $V_-$.
In Eq.(\ref{b1}), we express $B$ in terms of an equivalent set ($\tau_{\pm}$, $\epsilon$, $\eta$), which is more convenient when we want to take the thin-wall limit. Here, we 
also see how $B$ behaves as we move away from the triangular potential: the leading order correction is automatically incorporated into the two tension components $\tau_{\pm}$ by 
varying appropriately the 2/3 coefficient in Eq.(\ref{eq2}).

If $\lambda_-=\lambda_+$, we see that $\beta$ simplifies somewhat,
\begin{equation}
\beta= \frac{(1- \eta)}{2 \sqrt{\eta}}
\q
where $0 < \eta <1$ is given in Eq.(\ref{eta}); 
so the dividing point is at $\eta_c=(\sqrt{2} - 1)^2 = 0.17$. 
Often times in our analysis below, we shall restrict ourselves to this special case.

\section{Turning on Gravity}

Let us now turn on gravity \cite{Coleman:1980aw}, 
so the triangular potential has five parameters, namely ($\Delta V_{\pm}$, $\Delta \lambda_{\pm}$, $V_T$). As noted above, at times, it may be convenient to choose an alternative but equivalent set of parameters.

To emphasize the effect of gravity, let us consider the situation where 
\begin{equation}
V_T \gg \Delta V_-=V_T-V_-
\label{high energy}
\q
where $V_T/M_P^4 \ll 1$ so gravitational effects can be important for the tunneling while the semi-classical approximation is still valid. 

The equation (\ref{eqphi}) now reads,
\begin{equation}
r'^2=1+\frac{r^2}{3 M_p^2} (L-V_T),
\label{rho_1}
\end{equation}
where $L=\phi'^2/2 -V+V_T$ is the (shifted) Euclidean energy. By (\ref{high energy}), for large $V_T$, we use the approximation
\begin{equation}
r'^2\sim 1-\frac{r^2}{3 M_p^2}  V_T=1-H^2 r^2 .
\label{decoupled_rho}
\end{equation}
where 
\begin{equation}
H^2= V_T/3 M_p^2
\q 
is the Hubble constant of the de Sitter space with vacuum energy $V_T$. So we have 
\begin{equation}
r(\xi)=H^{-1} \sin(H \xi)
\label{Hubblesol}
\q
where $r(0)=r(\pi/H)=0$. That is, the range of $\xi$ is bounded, $0 \le \xi \le \pi/H$.

It is easy to check that this approximation is self-consistent, i.e., $\big| L(\xi) \big| \ll V_T$.
 For a tunneling solution, $\phi(\xi) \in [\phi_-,\phi_+]$, so 
\begin{equation}
L\geq 0.
\q
Using Eq.(\ref{eqphi}),
\begin{equation}
L'=\bigg(\frac{1}{2} \phi'^2-V(\phi)\bigg)'=-3\frac{r'}{r}\phi'^2 \sim -3 H \cot(H \xi) \phi'^2
\label{energy_conservation}
\end{equation}
When $\xi \in [0,\pi/2H)$, $L'\leq 0$ and if $ \xi \in (\pi/2H,\pi/H]$, $L'\geq 0$. Therefore $\max{L}=\max\{L(0),L(\pi/H)\}\leq \Delta V_-$. So $|L| \ll V_T$, and Eq.(\ref{Hubblesol}) is valid to the leading order in $\Delta V_-/V_T$,  For later purposes, we note that 
\begin{equation}
S_E(\phi_+) = -\frac{24 \pi^2 M_p^4}{V_+}
\label{Sphi+}
\q
  
The equations (\ref{eqphi}, \ref{eqr}) are now decoupled, and
\begin{equation}
\phi''+3H \cot(H \xi) \phi'=\frac{dV}{d\phi}.
\label{decoupled_phig}
\end{equation}
and can be solved with appropriate boundary conditions. This we shall do in the next section.

The bounce solution $\phi(\xi)$ will modify the geometric background used, i.e., $r(\xi)$ in (\ref{Hubblesol}). As we shall see, it is important to include this back-reaction. Inserting $\phi(\xi)$ back into Eq.(\ref{rho_1}), we get a first-order differential equation of the leading-order correction of $r$, namely $\delta r(\xi)$,
\begin{equation}
\cos(H\xi) \delta r'=- H \sin(H \xi) \delta r + \frac{L(\xi)}{2V_T} \sin^2(H \xi) . \label{drho}
\end{equation}
The formal solution for $\delta r$ is,
\begin{equation}
\delta r(\xi)=\cos(H\xi) \cdot \int_0^\xi d \eta \tan^2(H\eta) \frac{L(\eta)}{2 V_T},
\q
where the superficial singularity at $\eta=\pi/(2H)$ can be regularized as 
\m
\delta r(\xi)&=&\cos(H\xi) \cdot \int_0^\xi d \eta \tan^2(H\eta) \frac{L(\eta)-L(\frac{\pi}{2H})}{2 V_T}\nonumber \\ 
&+&  \frac{H^{-1} \sin(H \xi)-\xi \cos(H \xi)}{2 V_T} \cdot L(\frac{\pi}{2H}).
\label{delta_rho}
\n
Notice that by (\ref{energy_conservation}), $L'(\pi/2H)=0$ and the integral is well defined. Now we can consider the Euclidean action for this solution, to the leading order of $\Delta V_-/V_T$,
\m
S_E&=&4 \pi^2 \int_0^{\xi_{max}} d\xi \bigg((r+\delta r )^3 V(\xi)- 3(r+\delta r ) M_p^2 \bigg)\nonumber \\
&=& 4 \pi^2 \int_0^{\pi/H} d\xi \bigg(r^3 V_T- 3r M_p^2 \bigg) \label{action_vt} \\  
&+& 4 \pi^2 \int_0^{\pi/H} d\xi \bigg(\delta r (3r^2 V_T- 3 M_p^2)+r^3 \big(V(\xi)-V_T\big)\bigg) \label{action_leading} \\
&+& O\bigg(\frac{\Delta V_-^2}{V_T^2}\bigg) \label{action_higher}
\n
We separate the result into three parts. The first part (\ref{action_vt}) is $-24 \pi ^2  M_p^4/ V_T$, the de-Sitter space Euclidean action. The second part (\ref{action_leading}) is suppressed by a small factor $\Delta V_-/V_T$  comparing with the first part, because by (\ref{delta_rho}) the magnitude of $\delta r$ is of $H^{-1} \Delta V_-/V_T$, and the magnitude $V(\xi)-V_T$ is at most $\Delta V_-$. We will explicitly see that the two terms in the second part, which correspond to $\delta r$ and $\phi$ contribution are of the same order and therefore the computation of $\delta r(\xi)$ in Eq. (\ref{delta_rho}) is important. The last part represents all the high order terms of $\Delta V_-/V_T$. Notice that by the perturbation $\delta r$, $\xi_{max}$ is no longer $\pi/H$. However, the action change induced by $(\xi_{max}-\frac{\pi}{H})$ is of the second order of the $\Delta V_-/V_T$ and hence is included in (\ref{action_higher}). 

In conclusion, if we can get the analytic solution of (\ref{decoupled_phig}) then the leading order of $r$ is immediately given by the integral in (\ref{delta_rho}). It then 
follows that (\ref{action_vt}) and (\ref{action_leading}) give the leading order Euclidean action. Subtracting $S_E(\phi_+)$ given by(\ref{Sphi+}) from it, we get the factor $B$. 

\section{The Four Scenarios of Tunneling in de Sitter Space}

In general, Eq. (\ref{decoupled_phi}) is still too difficult to solve in a way that the physics is transparent. However,
for the triangle potential, we can solve (\ref{decoupled_phi}) analytically. For the different choices of the five parameters 
$V_T$, $\Delta \phi_+$, $\Delta \phi_-$, $\lambda_{\pm}$, we find here are four different kinds of Euclidean solutions for $\phi$ (shown in Figure \ref{4cases}) : 
\begin{itemize}
\item Case (I)  $\phi(\xi)$ reaches both $\phi_+$ and $\phi_-$. This happens when $\alpha \lambda/{H^2 \Delta \phi_+} \geq 1$ and 
$(\lambda/{H^2}) I\big(H^2 \Delta \phi_+/{\lambda}\big)\geq \Delta \phi_-$ where $\alpha \sim 0.4$ and $I(x)$ is a monotonic function shown in Figure \ref{function I}.
\item Case (II)  $\phi(\xi)$ reaches $\phi_+$ but not $\phi_-$. That is, $\phi(\xi)$ reaches $\phi_+$ and $\phi_{t-}$, where $\phi_T \le \phi_{t-} < \phi_-$. This happens when both $\alpha {\lambda}/{H^2 \Delta \phi_+} \geq 1$ and 
$({\lambda}/{H^2}) I\big({H^2 \Delta \phi_+}/{\lambda}\big)<\Delta \phi_-$ are satisfied.
\item Case (III). $\phi(\xi)$ reaches $\phi_-$ but not $\phi_+$. That is, $\phi(\xi)$ reaches $\phi_{f+}$, where $\phi_T \ge \phi_{f+} > \phi_+$. This case can happen only when the gradient towards the true vacuum is steeper than that towards the false vacuum.
\item Case (IV). $\phi(\xi)$ reaches neither $\phi_+$ nor $\phi_-$. That is, $\phi(\xi)$ reaches only $\phi_{f+}$ and $\phi_{t-}$. This case happens when ${\alpha \lambda}/{H^2\Delta \phi_+} <1$ is satisfied.

\end{itemize} 

In the absence of gravity, the triangle potential just has two kinds of bounce solutions \cite{Duncan:Jensen}, the cases (I) and (II). Here, including gravity, we have a new 
bounce solution, either case (III) or case (IV), because here there is a cut-off $\xi_{max}$ of the radius coordination $\xi$ due to the Hubble radius of de Sitter space. 
So in the finite range $0 \leq \xi \leq \pi/H$, the bounce solution may reach neither $\phi_+$ nor $\phi_-$. We will see that case (IV) is like HM tunneling, 
the case (I) is like thin-wall CDL tunneling, and case (II) is an intermediate case between the two limits. By adjusting the five parameters of the potential, 
we get a transition from HM tunneling to thin-wall CDL tunneling. 

For the triangular potential $V(\phi)$ shown in Figure \ref{Triangle}, we have, for large $V_T$,
\begin{equation}
\phi''+3H \cot(H \xi) \phi'=\frac{dV}{d\phi}= \pm \lambda_{\pm}
\label{decoupled_phi}
\q
whose general solution is given by
\begin{equation}
\phi(\xi) = a_1  \pm \frac{\lambda_{\pm}}{H^2}\left[ f_1(H\xi) + b_1 f_2(H\xi) \right]
\label{general_solution}
\q
where $f_1(H\xi)$ is the special solution and $f_2$ is the homogeneous solution.
\m 
f_1(x)&=&\frac{3}{8}  \cot^2 x - \frac{1}{24}  \csc^2 x - \frac{1}{3}  \ln (\sin x),\nonumber \\
f_2(x)&=&-\frac{1}{8} \csc^2\big(\frac{x}{2}\big) + \frac{1}{2} \ln \big(\tan(\frac{x}{2})\big) + \frac{1}{8}  \sec^2\big(\frac{x}{2}\big).
\label{f1f2}
\n 
Note that $f_2'(H\xi)= H/\sin^3 (H\xi)$.
The constants $a_1$ and $b_1$ are to be determined by the boundary and matching conditions.
We begin our discussion with case (IV) since it is new and its limiting case corresponds to HM tunneling. 

\begin{figure*}
  \centerline{
    \mbox{\includegraphics[width=1.95in]{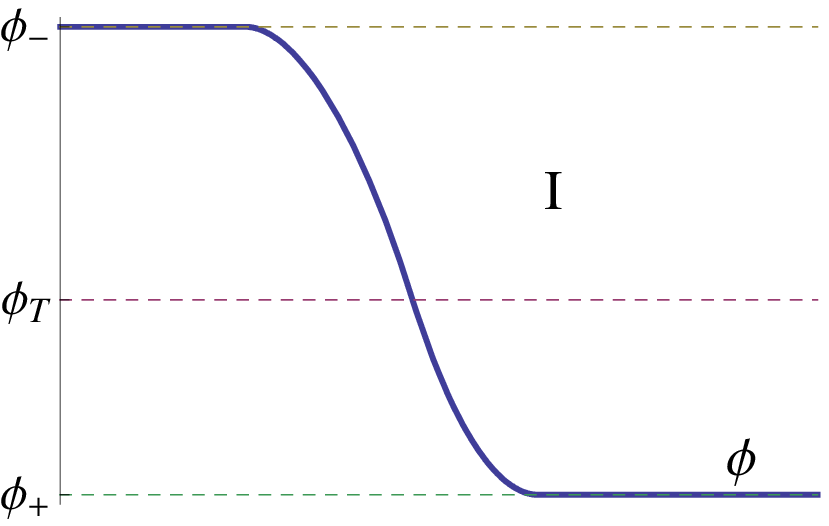}}
    \mbox{\includegraphics[width=1.95in]{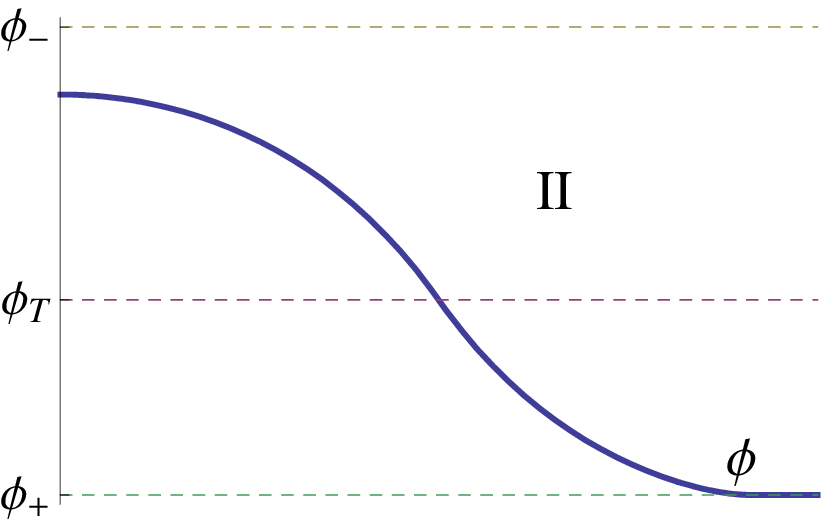}}
    }
   \centerline{
    \mbox{\includegraphics[width=1.95in]{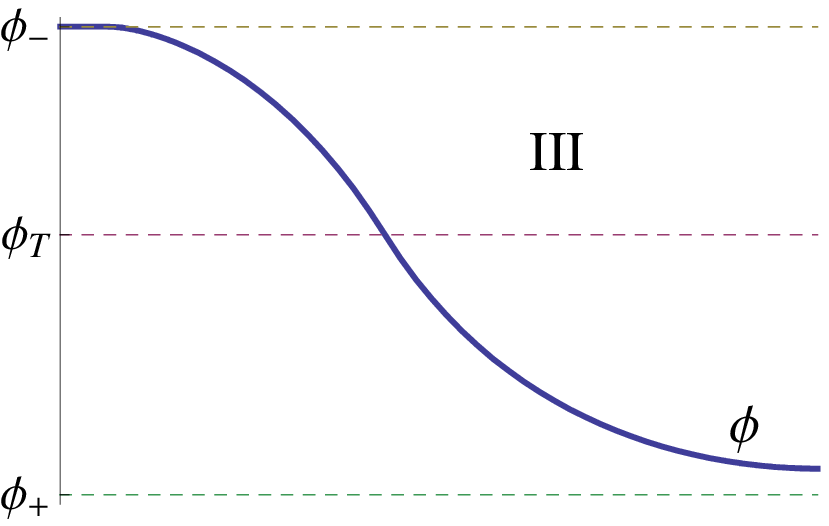}}
    \mbox{\includegraphics[width=1.95in]{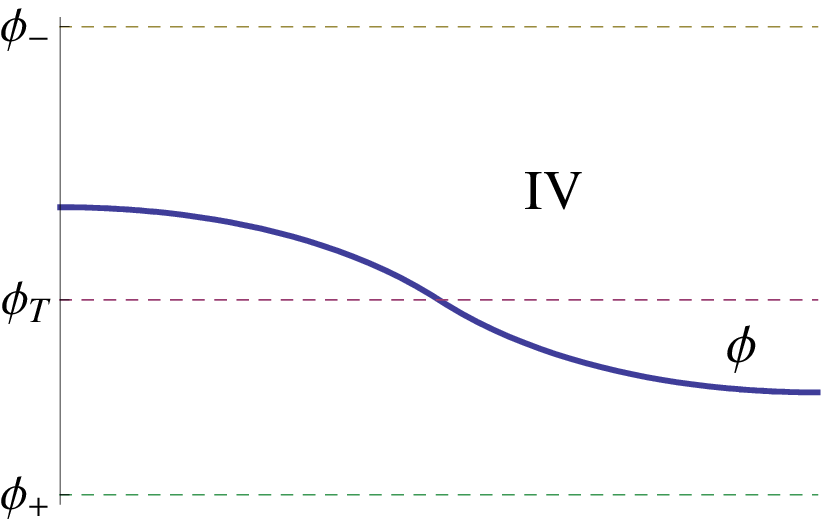}}
  }
  \caption{The Euclidean bounce solution of $\phi (\xi)$ as a function of $\xi = [0, \pi/H]$ in the four cases: (I) $\phi$ reaches the true vacuum at $\phi_-$ inside the nucleation bubble and the false vacuum at $\phi_+$ away from the outside of the bubble,  (II) $\phi$ reaches $\phi_+$ but not $\phi_-$, (III) $\phi$ reaches $\phi_-$ but not $\phi_+$ and (IV) $\phi$ reaches neither $\phi_-$ nor $\phi_+$; here the bounce starts at $\phi (0)=\phi_{t-} < \phi_-$ and ends at $\phi (\xi_{max}=\pi/H)=\phi_{f+} > \phi_+$. Case (I) and (II) can happen in the absence of gravity. Case (III) and (IV) can happen due to the presence of the de Sitter horizon. The thin-wall approximation is valid when the transition from $\phi_-$ to $\phi_+$ in case (I) is rapid. The HM formula is a good approximation when the variation of $\phi$ deviates little from $\phi_T$ in case (IV).
}
\label{4cases}
\end{figure*}

\subsection{Case (IV)}

For case (IV), the bounce solution of (\ref{decoupled_phi}) contains two pieces, which correspond to the $\phi_-$ side and the $\phi_+$ side,
\begin{equation}
\phi(\xi) = \left\{ \begin{array}{ll}
\phi_R(\xi) & \textrm{$\xi\in [0,\xi_{T}]$}\\
\phi_L(\xi) & \textrm{$\xi\in [\xi_{T},\pi/H]$}\\
\end{array} \right.
\label{case_4}
\end{equation} 
where $\phi_- >\phi_{t-} \geq \phi_R(\xi)\geq \phi_T$ and $\phi_+ < \phi_{f+} \leq \phi_L(\xi)\leq \phi_T$, whose general form can be obtained analytically. The boundary and matching conditions are
\begin{equation}
\left\{ \begin{array}{l}
 \textrm{$\phi'_R(0)=0$}\\
 \textrm{$\phi_R(\xi_{T})=\phi_L(\xi_{T})=\phi_T$}\\
 \textrm{$\phi_R'(\xi_{T})=\phi_L'(\xi_{T})$}\\
\textrm{$\phi_L'(\pi/H)=0$}
\end{array} \right.
\label{boundary_condition_4}
\end{equation} 
The first and last conditions require the specific combination $f_1+2f_2/3$ in the solutions (\ref{general_solution}).
Imposing the boundary conditions at $\xi_T$, we have,
\begin{eqnarray}
\phi_R(\xi)&=&\phi_T- \frac{\lambda_-}{H^2} \big(f(H\xi)-f(H\xi_T)\big) \nonumber
\\
\phi_L(\xi)&=&\phi_T+\frac{\lambda_+}{H^2} \big(f(\pi-H\xi)- f(\pi - H \xi_T)\big),
\label{case_IV_solution}
\end{eqnarray}
where 
\begin{equation}
f(x) = f_1(x) +\frac{2}{3} f_2(x) =  \frac{1}{24}\bigg(4 \sec\bigg(\frac{x}{2}\bigg)^2- 8 \ln \big(2 \cos\bigg(\frac{x}{2}\bigg)^2\big) -9 \bigg).
\label{fx}
\q
where we note that $f(\pi - x) = f_1(x) - 2f_2(x)/3$.
Matching the derivative at $\xi_T$ determines $\xi_T$,
\begin{equation}
c= \frac{\lambda_-}{\lambda_+} = \frac{f'(\pi - H \xi_T)}{f'(H\xi_T)}
\label{xiTmatch}
\q
so all parameters are now fixed.

To simplify the discussion, let us first consider the symmetric case $\lambda=\lambda_+=\lambda_-$, so the analytical solution has a symmetry about $\xi=\pi/(2H)$, and the solution is particularly 
simple: $\xi_{T}=\pi/(2H)$. It is easy to check that $\phi(\xi)$ is a monotonic decreasing function. For this case (IV), we require that $\phi(\xi)$ does not reach $\phi_+$ 
or $\phi_-$; in terms of the solution (\ref{case_IV_solution}),
\begin{equation}
\phi_{f+} = \phi_L(\pi/H)>\phi_+,
\q
Introducing a useful dimensionless parameter
\begin{equation}
 \gamma= \frac{\lambda}{H^2\Delta \phi_+} 
 \q
the above condition means 
\begin{equation}
\alpha \gamma = \frac{\alpha \lambda}{H^2\Delta \phi_+} <1
\label{case_IV_condition}
\end{equation}
where
\begin{equation}
\alpha= f(\pi/2)-f(0) = -\frac{1}{24} + \bigg(\frac{5}{24} +\frac{ \ln 2}{3}\bigg) \sim 0.398
\q
is a numerical constant. The similar condition $\phi_R(0)<\phi_-$ is satisfied automatically since $\Delta \phi_- > \Delta \phi_+$. 
In conclusion, if the parameters of the potential satisfy (\ref{case_IV_condition}), then the bounce solution is case (IV). 

Before computing $\delta r$ and $S_E(\phi)$, we estimate the magnitude of $\phi(\xi)$. By (\ref{case_IV_solution}), 
\begin{equation}
|\phi(\xi)-\phi_T|\sim \frac{\lambda}{H^2}.
\q
When $\lambda$ is small and $H$ is large, $\phi(\xi)$ is confined in a small region which centers at $\phi_T$. Because $\phi \equiv \phi_T$ is the Hawking-Moss bounce solution, 
case (IV) tunneling is like a fluctuation around HM tunneling and will approach the Hawking-Moss solution when $\lambda/H^2$ is small. 

Further, for the (shifted) Euclidean energy $L$, 
\begin{equation}
\frac{1}{2} \phi'^2\sim \frac{\lambda^2}{H^2}, \ |V-V_T| \sim \frac{\lambda^2}{H^2}
\q
Hence $L/V_T\sim \lambda^2 H^{-2}/ V_T$. By the condition (\ref{high energy}) and (\ref{case_IV_condition}), the ratio $\lambda^2/(H^2 V_T)\ll 1$ and hence $L/V_T\ll 1$ and the 
expansion (\ref{delta_rho}) and (\ref{action_leading}) works for case (IV). The only new issue is that for case (IV), the small expansion factor is $\lambda^2/ H^{2} V_T$ not 
$\Delta V_-/V_T$, simply because the bounce solution (\ref{case_IV_solution}) does not ``feel" $\phi_-$. The integral $(\ref{delta_rho})$ is carried out analytically, and by 
(\ref{action_vt}),(\ref{action_leading}) and (\ref{action_higher}), we find that 
the two terms in the leading order correction have the same form,
\begin{equation}
S_E(\phi) = \frac{24 \pi ^2 M_p^4}{V_T} - C \frac{\lambda^2 M_p^6}{V_T^3} + O\bigg(\frac{\lambda^4 M_p^8}{ V_T^5}\bigg).
\label{SEHMB}
\q
where $C=C_1+C_2\sim 65.49$ is a positive constant. $C_1\sim -196.47$ corresponds to the $\delta r(\xi)$ contribution and $C_2\sim 261.96$ is from $\phi(\xi)$. So we see explicitly that the two terms in Eq. (\ref{action_leading}) are comparable. That is, the back-reaction is important. The above result has a simple interpretation that we shall describe in Sec. 6. 

Since the Hawking-Moss scenario is a special limit in this case, let us treat this decay as a HM transition, with a correction to the HM formula that can be explicitly evaluated.
So $B$ is given by
\begin{eqnarray}
B=S_E(\phi) -S(\phi_+)
&=& 24 \pi ^2  M_p^4 \bigg( \frac{1}{V_+} - \frac{1}{V_T} \bigg) - C \frac{\lambda^2 M_p^6}{V_T^3} + O\bigg(\frac{\lambda^4 M_p^8}{ V_T^5}\bigg).
\label{B_case_IV}
\end{eqnarray}
The first term is just the Hawking-Moss  bounce $B_{HM}$ (\ref{BHM0}). 
while the second term, correction from this bounce solution, lowers $B$ and therefore the actual tunneling rate is larger than that given by the HM formula.

For $V_T \gtrsim V_+$, the first term in the bounce (\ref{SEHMB}) is largely cancelled by $S(\phi_+)$, so the corrections can be important. Rewriting the bounce formula as
\begin{equation}
B= B_{HM} \left( 1 - \frac{C}{72 \pi^2}\gamma \bigg(\frac{V_+}{V_T}\bigg) + . . . \right) = B_{HM} \left( 1 - 0.2317 (\alpha \gamma) \bigg(\frac{V_+}{V_T}\bigg)+ . . . .\right)
\q
Since both $\alpha \gamma < 1$ (\ref{case_IV_condition}) and $V_+/V_T < 1$, we see that the correction to the HM formula is at most $23\%$. \footnote{Notice that the perturbation series (\ref{action_leading}), (\ref{action_leading}) and (\ref{action_higher}) start from the de-Sitter space Euclidean action, not the HM tunneling exponential factor $B_{HM}$. So the correction, though large  compared to $B_{HM}$, is still much smaller than the zero order, de-Sitter space Euclidean action. Hence the leading correction yields a very good approximation here. That is, the last term in Eq.(\ref{SEHMB}) or (\ref{B_case_IV}) is negligible.}


The picture of case (IV) is consistent with the interpretation proposed in \cite{Goncharov}, \cite{Starobinsky}, \cite{GLM}. HM tunneling should be interpreted as a quantum
fluctuation up the potential barrier. Here we see that the field will thermally fluctuate from $V_+$ at $\phi_+$ up to $V(\phi_L(\pi/H))< V_T$ at $\phi_L(\pi/H) > \phi_+$, tunnel to $\phi_R(0) < \phi_-$ and then classically roll down to $\phi_-$.
In the limit where $\gamma \Delta V_+/V_T \rightarrow 0$, $V_T$ is simply reached via the thermal fluctuations alone.
Here, we see that the correction is typically not very big and the HM formula is quite good in general.
On the other hand, as $\gamma \rightarrow 1/\alpha=2.51$, $\phi_L(\pi/H) \rightarrow  \phi_+$; the outside of the bubble can reach the false vacuum within the horizon. When this happens, the HM formula is no longer accurate.

\begin{figure}[t]
\centering
\includegraphics{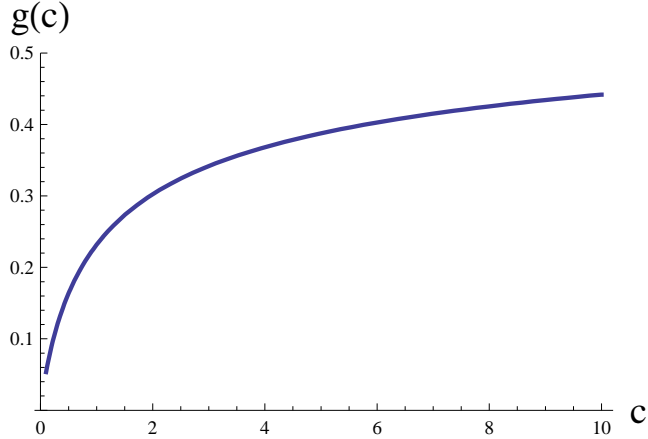}
\caption{The function g(c).}
\label{function_g}
\end{figure}

Now let us consider the more general case where $c=\lambda_-/\lambda_+\not=1$. The solution of $\xi_T$ should be determined by 
\begin{equation}
c \bigg(-f_1'(\xi_T) - \frac{2}{3} f_2'(\xi_T)\bigg) = f_1'(\xi_T) - \frac{2}{3} f_2'(\xi_T)
\q
which can be solved numerically. For large $c$, the solution of $\xi_T$ is determined by series expansion, $\xi_T\sim \sqrt[4]{16/(3 c)}$. The condition of the case (IV) is now
\begin{equation}
\alpha(c) \gamma <1,
\q
where $\alpha(c)=\big(f_1(\xi_T) - \frac{2}{3} f_2(\xi_T)\big)-\big(f_1(\pi) - \frac{2}{3} f_2(\pi)\big)$, where $\alpha(1)=\alpha\sim 0.398$. For large $c$, by the asymptotic form of $\xi_T$, $\alpha(c)\sim \sqrt{C(c)/12}$. We can integrate the solution to get $\delta r$ and $B$, 
\m
B&=& 24 \pi^2 M_p^4 \bigg( \frac{1}{V_+} - \frac{1}{V_T} \bigg)- C(c) \frac{\lambda_+^2 M_P^6}{V_T^3} + O\bigg(\frac{\lambda^4 M_P^8}{ V_T^5}\bigg), \\
&=& B_{HM}\bigg(1-\frac{C(c)}{72 \pi^2} \gamma \frac{V_+}{V_T}\bigg)=B_{HM}\bigg(1-\frac{C(c)}{72 \pi^2 \alpha(c) } \alpha(c) \gamma \frac{V_+}{V_T}\bigg)\nonumber \\
&\equiv& B_{HM}\bigg(1-g(c) \big(\alpha(c) \gamma\big) \frac{V_+}{V_T}\bigg).
\n
As before, both $\alpha(c) \gamma<1$ and $V_+/V_T<1$. The function $g(c)$ is plotted in Figure \ref{function_g},
When $C$ is large, $C(c)$ grows as $\sqrt{c}$ and numerically the proportional coefficient is $\sim  131$. Hence, by the asymptotic form of $\alpha(c)$,
\begin{equation}
g(c) \to 0.64, \quad \quad  \mbox{when} \quad  c\to \infty.
\q
which means that the leading order correction is at most about $64\%$. We see that the HM formula is an over estimate of the value of $B$, by as much as a factor of three.

\subsection{Case (III)}

Here we shall simply show that this case, where $\phi(\xi)$ reaches $\phi_-$ but not $\phi_+$ in the Euclidean solution,
exists for some choice of the potential.
Following the above solution (\ref{case_IV_solution}), we have 
\begin{eqnarray}
\Delta \phi_R = \phi_R(0) - \phi_T &=& \frac{\lambda_-}{ H^2}  \big(f(H\xi_T)- f(0)\big) \le \Delta \phi_-\nonumber \\
\Delta \phi_L = \phi_T - \phi_L(\pi/H) &=&\frac{\lambda_+}{H^2} \big(f(\pi-H\xi_T)- f(\pi)\big) \le \Delta \phi_+
\label{case3}
\end{eqnarray}
Let $\Delta V_R= V_T-V(\phi_R(0)) \le \Delta V_-$ and $\Delta V_L= V_T-V(\phi_L(\pi/H)) \le \Delta V_+$. Case (IV) corresponds to
\begin{equation}
 \Delta V_R < \Delta V_-,  \quad \quad \Delta V_L < \Delta V_+
\q
while case (III) would be reached if
\begin{equation}
 \Delta V_R = \Delta V_-, \quad \quad \Delta V_L < \Delta V_+.
 \q
Using Eq.(\ref{fx}), Eq.(\ref{xiTmatch}) can be rewritten as 
\begin{equation}
c= \frac{\lambda_-}{\lambda_+} =\frac{2\cos^2(H\xi_T/2) + \cot^2 (H\xi_T/2)}{2\sin^2(H\xi_T/2) + \tan^2 (H\xi_T/2)}
\label{HxiTy}
\q
Moving slightly away from the symmetric case, we have, from the above equation (\ref{HxiTy}), for small deviations,
\begin{equation}
H\xi_T \simeq \pi/2 -\frac{1}{3}(c-1).
\q
so we have
\begin{equation}
\frac{\Delta V_R}{\Delta V_L} = \frac{c\Delta \phi_R}{\Delta \phi_L}  \simeq c^2\frac{9\alpha  - 2 (c-1)}{9\alpha+ 2 (c-1)}\simeq 1+(2-\frac{4}{9\alpha})(c-1)
\q
Hence for $c>1$, we have 
\begin{equation}
\Delta V_R >  \Delta V_L 
\q 
For $\Delta V_- > \Delta V_R$ and $\Delta V_+ > \Delta V_L$, we have still have the case (IV). Now we can consider potentials where 
\begin{equation}
\Delta V_R \ge \Delta V_- >  \Delta V_+ > \Delta V_L 
\label{case234}
\q
(decreasing  $\Delta \phi_-$ while keeping other parameters fixed) which belongs to case (III). This demonstrates that case (III) exists for some types of potential. Here is an explicit example: when $\Delta \phi_+=0.005$, $\Delta \phi_-=0.004$,
$\lambda_-=1\times 10^{-7}$, $\lambda_+=0.5\times 10^{-7}, V_T=0.77 \times
10^{-6}$, the bounce solution, which is plotted in Figure.\ref{4cases}, does reach $\phi_-$ but not $\phi_+$. 

On the other hand, for $c<1$ (that is, when the gradient towards the false vacuum is steeper than that towards the true vacuum), the Euclidean solution has  
$\Delta V_R < \Delta V_L$. This means the tunneling is from $V_L=V_T-\Delta V_L$ going up to $V_R=V_T-\Delta V_R$. As noted earlier, this tunneling up phenomenon does not 
happen in the absence of gravity. So this possibility is a gravitational effect.  This phenomenon has been studied in \cite{LW}.

Note that this case which reaches $\phi_-$ but not $\phi_+$ does not exist for the symmetric ($\lambda=\lambda_+=\lambda_-$) case. 
In this case, $c=1$ so that $\Delta V_R = \Delta V_L$ and the above condition (\ref{case234}) cannot be satisfied unless  $\Delta V_R = \Delta V_+ = \Delta V_L$, which means that $\phi$ does reach the false vacuum value. This corresponds to the cases we shall now turn to in the next subsection.

\subsection{Case (II)}

To see the existence of this case, we may restrict ourselves to the symmetric ($\lambda_+=\lambda_-$) case. Since $\Delta \phi_- > \Delta \phi_+$we can have the situation where   
\begin{equation}
\Delta \phi_- > \frac{\alpha \lambda}{H^2} \ge \Delta \phi_+
\q
That is, the condition (\ref{case_IV_condition}) does not hold. 
Instead, we now have
\begin{equation}
\alpha \gamma = \alpha \frac{\lambda}{H^2 \Delta \phi_+} \geq 1.
\label{case_II_condition_1}
\end{equation}

In this case, $\phi(\xi)$ reaches the false vacuum $V_+$ at $\phi_+$ outside the bubble within the horizon but the bubble inside never reaches the true vacuum $V_-$ at $\phi_-$ at the moment of creation. In this case, the solution of (\ref{decoupled_phi}) contains three pieces.
\begin{equation}
\phi(\xi) = \left\{ \begin{array}{ll}
\phi_1(\xi) & \textrm{$x\in [0,\xi_T]$}\\
\phi_2(\xi) & \textrm{$x\in [\xi_T,\xi_+]$}\\
\phi_+ & \textrm{$x\in [\xi_+,\pi/H]$}
\end{array} \right.
\label{case_2}
\end{equation} 
where $\phi_1(\xi)\geq \phi_T$ and $\phi_2(\xi)\leq \phi_T$. The general solution for $\phi_1(\xi)$ and $\phi_2(\xi)$, which satisfies $\phi_1'(0)=0$, is 
\m 
\phi_1(\xi)=\frac{\lambda}{H^2} \bigg(C_1 -f_1(H \xi)-\frac{2}{3} f_2(H \xi) \bigg)  \nonumber \\
\phi_2(\xi)=\frac{\lambda}{H^2} \bigg(C_2 +f_1(H \xi)+A f_2(H \xi) \bigg),
\n 
where $f_1$ and $f_2$ are given by Eq.(\ref{f1f2}).
The remaining boundary and matching conditions are,
\begin{equation}
\left\{ \begin{array}{l}
 \textrm{$\phi_1(\xi_T)=\phi_2(\xi_T)=\phi_T$}\\
 \textrm{$\phi_1'(\xi_T)=\phi_2'(\xi_T)$}\\
\textrm{$\phi_2(\xi_+)=\phi_+$}, \textrm{ $\phi_2'(\xi_+)=0.$} \\
\end{array} \right.
\label{boundary_condition_2}
\end{equation} 
We have five parameters to determine, $C_1,C_2,A,\xi_T,\xi_+$, and also five equation from (\ref{boundary_condition_2}). It is helpful to rewrite (\ref{boundary_condition_2}) as three equations only in $A, \xi_T,\xi_+$
\begin{equation}
\left\{ \begin{array}{l}
 \textrm{$f_1(H \xi_T) - f_1(H \xi_+) + A \big(f_2(H \xi_T) - f_2(H \xi_+)\big) = H^2\Delta \phi_+ /\lambda =1/\gamma$}\\
 \textrm{$f_1'(H \xi_T) + (A/2 +1/3) f_2'(H \xi_T)  = 0$}\\
\textrm{$f_1'(H \xi_+) + A f_2'(H \xi_+) = 0$} 
\end{array} \right.
\label{boundary_condition_2_reduced}
\end{equation}
Because the functions $f_1$ and $f_2$ contain only pure number coefficients, the dimensionless parameters $A,H \xi_T,H \xi_+$ are determined just by the combination $\gamma= \lambda/H^2 \Delta \phi_+$. So we may rewrite them as functions of $1/\gamma$ which can be obtained numerically. Then we can insert $A,H \xi_T,H \xi_+$ back into (\ref{boundary_condition_2}), we will get all the parameters. 

For case (II), $\phi(\xi)$ does not reach $\phi_-$, or
\begin{equation}
\phi_1(0)-\phi_T < \Delta \phi_-,
\q
or
\begin{equation}
 f_1(H\xi_T)+\frac{2}{3} f_2(H\xi_T)-f_1(0)-\frac{2}{3} f_2(0) \equiv I(1/\gamma) <  \frac{ H^2 \Delta \phi_-}{\lambda}.
\q
Notice that the left hand side is just a function of $1/\gamma=H^2 \Delta \phi_+/\lambda$, which we call $I(1/\lambda)$. So we need,
\m 
\frac{\lambda}{H^2} I\bigg(\frac{H^2 \Delta \phi_+}{\lambda}\bigg)<\Delta \phi_-.
\label{case_II_condition_2}
\n 
Equation (\ref{case_II_condition_2}), together with (\ref{case_II_condition_1}), is the condition for case (II). The function $I(x)$ is a monotonic function which is plotted in Figure.\ref{function I}. Notice that $I(x)>x$, therefore,
\begin{equation}
\phi(0)-\phi_T>\Delta \phi_+
\q 
which means the quantum tunneling is from the false vacuum $V_+$ to someplace lower than $V_+$ but not exactly at $V_-$. After the realization of quantum tunneling, $V$ will continue to drop classically until reaching $V_-$, the true vacuum. 
\begin{figure}[t]
\centering
\includegraphics{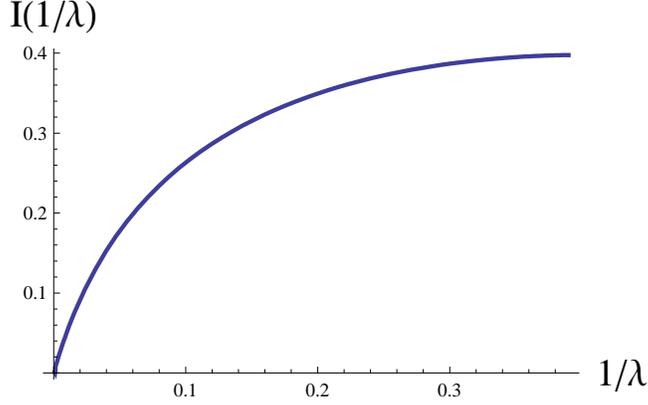}
\caption{The function $I(1/\gamma)$. Note that $1/\gamma < \alpha =0.398$. }
\label{function I}
\end{figure}
After getting the solution $\phi$, one uses the formulae (\ref{delta_rho}), (\ref{action_vt}) and (\ref{action_leading}) to get the factor $B$ for the case (II). Notice here $\phi$ does not reach $\phi_-$, so actually, the small expansion factor is not $\Delta V_-/V_T$ but an even smaller factor $\Delta V_+/V_T$.

\subsection{Case (I)}

In the this case, $\phi(\xi)$ reaches both $\phi_+$ and $\phi_-$. Therefore the condition is 
\m 
\alpha \gamma= \alpha \frac{\lambda}{H^2 \Delta \phi_+} &\geq& 1 \nonumber 
\\
\frac{\lambda}{H^2} I\bigg(\frac{H^2 \Delta \phi_+}{\lambda}\bigg)&\geq &\Delta \phi_-
\label{case_I_condition}.
\n 
$\phi(\xi)$ contains four pieces, 

\begin{equation}
\phi(\xi) = \left\{ \begin{array}{ll}
\phi_- & \textrm{$x\in [0,\xi_-]$} \\
\phi_1(\xi) & \textrm{$x\in [\xi_-,\xi_T]$}\\
\phi_2(\xi) & \textrm{$x\in [\xi_T,\xi_+]$}\\
\phi_+ & \textrm{$x\in [\xi_+,\pi/H]$}
\end{array} \right.
\label{case_1}
\end{equation}
where $\phi_1(\xi)\geq \phi_T$ and $\phi_2(\xi)\leq \phi_T$. The general solution for $\phi_1(\xi)$ and $\phi_2(\xi)$ is,
\m 
\phi_1(\xi)=\frac{\lambda}{H^2} \bigg(C_1 -f_1(H \xi)+A_1 f_2(H \xi) \bigg)  \nonumber \\
\phi_2(\xi)=\frac{\lambda}{H^2} \bigg(C_2 +f_1(H \xi)+A_2 f_2(H \xi) \bigg),
\n 
where $f_1$ and $f_2$ are defined in (\ref{f1f2}). The boundary conditions are,
\begin{equation}
\left\{ \begin{array}{l}
\textrm{$\phi_1(\xi_-)=\phi_-$} , \textrm{ $\phi_1'(\xi_-)=0$} 
\\
 \textrm{$\phi_1(\xi_T)=\phi_2(\xi_T)=\phi_T$}\\
 \textrm{$\phi_1'(\xi_T)=\phi_2'(\xi_T)$}\\
\textrm{$\phi_2(\xi_+)=\phi_+$}, \textrm{ $\phi_2'(\xi_+)=0.$}
\end{array} \right.
\label{boundary_condition_1}
\end{equation} 
So we have seven parameters to determine, $C_1,C_2,A_1,A_2,\xi_-,\xi_T,\xi_+$, and also seven equations from (\ref{boundary_condition_1}). Like case (II), we can reduce the number
of both the parameters and the equations and then solve it. The solution looks like the thin-wall tunneling solution in \cite{Coleman:1980aw}. The reason is, by the condition
(\ref{case_I_condition}), $\Delta \phi_+$ cannot be very small comparing with $\Delta \phi_-$, so the energy difference of the two vacua $\epsilon=V_+-V_-$ cannot be large and the
thin-wall approximation may apply.
Again, after getting the solution of $\phi(\xi)$, we can get the Euclidean action and $B$ by (\ref{delta_rho}), (\ref{action_vt}) and (\ref{action_leading}).

It is clear that the thin-wall approximation belongs to case (I), since the inside of the bubble reaches $\phi_-$ while the outside reaches $\phi_+$ (see Figure \ref{4cases}). 
The thin-wall approximation requires a rapid transition of $\phi$ from $\phi_-$ to $\phi_+$.
This case has been analyzed in Ref.\cite{Huang:}, \cite{Lee}, \cite{Parke}. For the sake of completeness, let us review the basic result following \cite{Huang:}.
 In the thin-wall approximation ($\epsilon \rightarrow 0$), we may divide the integration for the bounce $B$ into
three parts. Outside the bubble, $\phi=\phi_+$ and thus \begin{equation}
B_{out}=0. \q In the wall, we have \begin{equation} B_{wall}=2\pi^2r^3\tau, \q
where $r$ is the bubble size and $\tau$ is the tension of the wall
which is decided by the barrier between the false and true vacua,
\begin{equation}
\label{tension}
\tau \simeq \int^{\phi_+}_{\phi_-} d \phi \sqrt{2[V(\phi)-V(\phi_+)]}
\q
Inside the bubble, $\phi=\phi_-$ is a
constant and Eq.(\ref{eqr}) becomes 
\begin{equation}
\label{echi}
 d\xi=dr(1-\kappa r^2V/3)^{-1/2}.
 \q 
 Hence
\begin{equation} 
S_{E,in}(\phi)=-{12\pi^2\over
\kappa}\int_0^r {\tilde r}d{\tilde r}(1-\kappa V(\phi) {\tilde
r}^2/3)^{1/2}. 
\q
Summing the three parts of $B$, we obtain 
\begin{equation}
 B=2\pi^2r^3\tau+{12\pi^2\over \kappa^2}
\[{1\over V_-}\(\(1-\kappa r^2V_-/3\)^{3/2}-1\)-{1\over V_+}\(\(1-\kappa r^2V_+/3\)^{3/2}-1\)\]. \label{gb}
\q 
The coefficient $B$ is stationary at $r=R$ which satisfies
\begin{equation}
 {1\over R^2}={\epsilon^2\over 9\tau^2}+{\kappa(V_++V_-)\over
6}+{\kappa^2\tau^2\over 16}, \label{sob}\q 
where $\epsilon=V_+-V_-$.
So
\begin{equation}
B_{tw}=2\pi^2R^3\tau+{4\pi^2\over \kappa}
\[R_-^2\(\(1- H_-^2R^2\)^{3/2}-1\) - R_+^2\(\(1-H_+^2R^2\)^{3/2}-1\)\] \label{gbsoln}
\q 
where $H_{\pm}^{-1} = R_{\pm}=(\kappa V_{\pm}/3)^{-1/2}$ and the subscript ``tw'' means the thin-wall approximation. 
Note that the last term is proportional to $\kappa^2$ and so is very small most of the time.
According to Eq.(\ref{sob}), we can easily check that the bubble
radius at the moment of materialization is not larger than the event
horizon $R_+$ of the de Sitter space in false vacuum.
This is reasonable; otherwise the bubble cannot be
generated causally. 

For the special case with $R\ll R_+$, the bubble size is much
smaller than the curvature radius of the background and gravity does
not play a big role. In this limit $B$ becomes \begin{equation}
B=2\pi^2r^3\tau-{\pi^2\over 2}r^4\epsilon.\label{vb}\q 
Here, $B$ is stationary at \begin{equation} r=R_0={3\tau\over \epsilon}, \label{ssob}\q
so  
\begin{equation}
 B_{tw}\sim {27\pi^2\over 2}{\tau^4\over
\epsilon^3}.\q

In the other limit of $R\simeq R_+$ which
corresponds to $V_+\simeq V_-=V\gg V_s={2\epsilon^2/
3\kappa\tau^2}+{3\kappa\tau^2/8}$. This happens at the high
energy scale in the landscape, and the bubble radius is
given by $R\simeq \sqrt{3/ \kappa V}$. 
Now $B$ is dominated
by the first term in Eq.(\ref{gb}), namely 
\begin{equation}
\label{BLR}
 B_{tw}\simeq 6\sqrt{3}\pi^2\tau (\kappa V)^{-3/2} = {2{\pi^2 \tau} \over H^3} 
 \q 
In the units where $M_p=1$, or
equivalently $\kappa=1$, the tension of the bubble satisfies $\tau
\ll 1$. In Planck region $(V\sim 1)$, $B\ll 1$ and $\Gamma\sim 1$.
At low energy scale, the background curvature radius is quite large
and the bubble size is relatively small, and then the tunneling rate
is insensitive to the vacuum energy of the false vacuum.

\begin{figure}[t]
\centering
\includegraphics[width=4.0in]{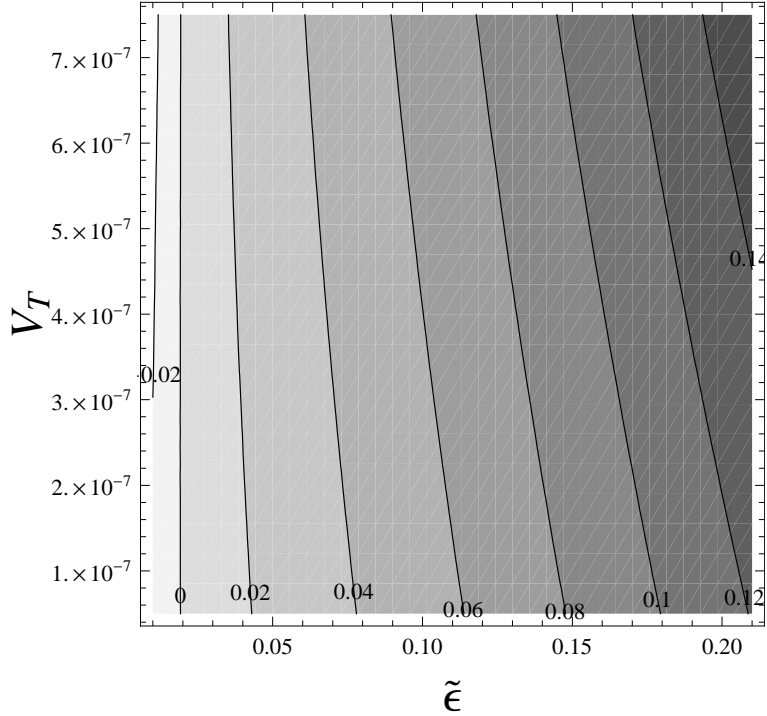}
\caption{The contour plot of $(B_{tw}-B)/B$ as the function of $\tilde \epsilon$ and $V_T$. The horizontal axis is $\tilde \epsilon$ and the vertical axis is $V_T$. The contour line labels the value $(B_{tw}-B)/B$ and the darker regions have larger differences.}
\label{Case_I_thin_wall}
\end{figure}

The familiar CDL thin-wall approximation occurs in case (I), 
however, a solution of the case (I) is not
necessarily the ``thin-wall'' solution, i.e., the tunneling factor $B_{tw}$ determined by the thin-wall approximation may not be very accurate. To illustrate this point we compare the values of $B_{tw}$
with our computation for case (I).

We fix $G=1$, $\lambda=1\times 10^{-7}$, $\phi_T=0$, $\phi_+=-0.005$ and vary the variables $\phi_-$ and $V_T$.  We define
\begin{equation}
\tilde \epsilon=\frac{\epsilon}{\Delta V_+}=\frac{V_+-V_-}{V_T-V_+}=\frac{\phi_--|\phi_+|}{|\phi_+|}
\q 
which is always positive and we may expect that when $\tilde \epsilon$ is close to $0$, the thin-wall approximation is accurate. We reserve $B$ for our result for the tunneling exponential
factor from Eq.(\ref{delta_rho}), Eq.(\ref{action_vt}) and Eq.(\ref{action_leading}) while $B_{tw}$ denotes the counterpart of the thin-wall approximation formula (\ref{gbsoln}) given in Ref.\cite{Huang:}.  We plot the relative deviation of the two methods, $(B_{tw}-B)/B$, as a
function of $r$ and $V_T$ in the contours of Figure \ref{Case_I_thin_wall}.

In the contour figure we draw the range $0.01<\tilde \epsilon<0.2$ and $5 \times 10^{-8}<V_T<7.5 \times 10^{-7}$. The reason for the $V_T$ range choice is that if $V_T$ is too small then our
expansion is not good and if $V_T$ is too large, case (I) will turn into case (II).  From the figure we can see
\begin{itemize}
\item The contour lines are roughly vertical and the relative difference $(B_{tw}-B)/B$ increases from the left to right. It means that when $\tilde \epsilon$ is small the thin-wall
approximation coincides with our computation and so verifies that the thin-wall appoximation is good when $\epsilon$ is small.
\item However, the contour lines are not completely vertical but tilt from the top left to the bottom right. It means when $\tilde \epsilon$ is fixed and $V_T$ is increasing, the difference
is getting larger and larger. An immediate explanation is: in this situation the solution is moving toward case (II) and finally when tunneling becomes case (II) the difference is
very large, since the solution does not reach $\phi_-$ in contrast to the thin-wall approximation. Or by the Euclidean equation of motion, if $V_T$ is large, $H$, which serves as a
damping term, is also large and slows the evolution of the Euclidean solution. So the wall is not ``thin" as before.
 \end{itemize}

Although unlike case (IV), it is hard to get the analytic form of $B$ for case (I) because of the transcendental Eq.(\ref{boundary_condition_1}). However, it is helpful to look at the numerical computation. For example, for a potential with small $\epsilon=V_+-V_-$, like $\lambda=1\times 10^{-7}$, $V_T = 0.2\times 10^{-6}$, $\Delta \phi_+ = 0.005$ and $\Delta \phi_- = 0.0055$, the zero-order of the $S_E$, by Eq.(\ref{action_vt}), is $S_{E,0}=-1.87500000\times 10^6$. 
Eq.(\ref{action_leading}) gives the leading order correction $S_{E,1}=S_{E,1,\delta r}+S_{E,1,\phi}$, where $S_{E,1,\delta r}=4787.05$ is the first term in (\ref{action_leading}) from $\delta r$ contribution and $S_{E,1,\phi }=-7691.97$ is the second term induced by $\phi$. Therefore, we explicitly see that the leading-order corrections $S_{E,1,\delta r}$ and $S_{E,1,\phi}$ are much smaller than the zero order $S_{E,0}$ and the perturbation series for $S_E$ is valid. To get the factor $B$, we needs to compute $S(\phi_+)$ which is $-1.8796992 \times 10^6$, so
\begin{eqnarray*}
B&=&S_E -S(\phi_+)\sim S_{E,0}+S_{E,1,\delta r}+S_{E,1,\phi}-S(\phi_+)\\
&=&1794.32
\end{eqnarray*}
which is quite close to the thin-wall approximation result $B_{tw}=1848.76$ by Eq.(\ref{gb}). 

We see that it is crucial to include the back-reaction effect due to $\delta r$ here. In fact, if we drop the contribution from $\delta r$ in Eq.(\ref{action_leading}), then $B({\mbox{$\delta r$-excluded}) }\sim S_{E,0}+S_{E,1,\phi}-S(\phi_+) 
\sim -2992$,
which is negative and has no physical meaning. 
The importance of the $\delta r$ term in Eq.(\ref{action_leading}) cannot be over-emphasized.

\section{Discussion}

Let us comment on the physical meaning of the cases other than case (I). For case (II), after the realization of quantum tunneling, the field will evolve by the classical equation. Coleman \cite{Coleman:1977py} points out that, instead of going back to solve the Minkowski (Lorentz) equation, we can get the classical solution directly from the Euclidean solution by analytical continuation. For \cite{Duncan:Jensen}, we can verify that, after the quantum tunneling, the field will continue to roll down until reaching the true vacuum.  
This also happens in de Sitter space.
After the creation of the nucleation bubble with radius $R$ at time $t=0$, the evolution of the surface of the bubble is described by the Lorentzian action, where $\xi^2=R^2 \rightarrow {\bf r}^2 - t^2$. The center of the bubble starts at
$\xi=0$ and as $t$ increases, $\xi^2=-t^2$ becomes negative at the center. By analytic continuation, starting with  $\phi(r=0,t=0) < \phi_-$, $\phi(0,t)$ falls towards $\phi_-$. It may
overshoot and oscillates about the true vacuum at $\phi_-$. With Hubble (or any additional) damping, it will eventually settle at the true vacuum as the nucleation bubble
continues to grow. In case (III) and case (IV),  $\phi$ at the bubble outside is also expected to follow classical motion and roll down towards the false vacuum at $\phi_+$ (while in a GH thermal bath). However, since the bubble is growing rapidly at the same time, the true vacuum may be reached before $\phi$ has time to roll to the false vacuum.  

Let us show the different features of the tunneling by an explicit example. The type of tunneling is determined by the parameters of the potential, $V_T$, $\lambda_+$, $\lambda_-$, $\delta \phi_+$ and  $\delta \phi_-$. First, we want to see the dependence of the exponent $B$ on $H$, for large $H$.
We set $G=1$ and fix $\phi_T=0$, $\Delta \phi_-=0.007$, $\Delta \phi_+=0.005$, $\lambda_{\pm}=1\times 10^{-7}$ so the shape of the potential is fixed. We can move the potential up and down by varying $V_T$ (hence $H$) to see the dependence of $B$ on $H$. 
\begin{itemize}
\item The condition for case (IV), 
$\alpha \gamma < 1$ reads $V_T> 9.49476 \times 10^{-7}$. For example, we can plot the solution of $\phi(r)$ for $V_T=2 \times 10^{-6}$ in the last picture of Figure \ref{4cases}
and it is clear that neither $\phi_+$ nor $\phi_-$ is reached. The exponential factor $B$ can be obtained by Eq.(\ref{B_case_IV}), $B\sim 41.7301$. The HM tunneling exponential
factor for this potential is $B_{HM}\sim 46.8867$ hence the case (V) tunneling rate is close to the HM tunneling but faster than it. The relative difference of $B$ is about $11\%$.

\item When $\alpha \gamma \geq 1$, we need to consider the condition
\m
\frac{\lambda}{H^2} I\bigg(\frac{H^2 \Delta \phi_+}{\lambda}\bigg)<\Delta \phi_-.
\n
which reads $9.49476 \times 10^{-7}\geq V_T> 6.46586 \times 10^{-7}$. If it is satisfied, the tunneling is of case (II). For example, we plot the solution for $V_T=9.0 \times 10^{-7}$ in Figure \ref{4cases} and find that $\phi$ reaches $\phi_+$ but not $\phi_-$. 

\item If this condition is also violated, i.e., $V_T\leq 6.46586 \times 10^{-7}$,
the tunneling is case (I). \footnote{ Notice that our approximation works for 
$V_T \gg \Delta V_-$,
which reads $V_T \gg 7\times 10^{-10}$. } For example, the solution of $V_T=3.0 \times 10^{-7}$ is plotted in Figure \ref{4cases} and $\phi$ reaches both $\phi_+$ and $\phi_-$. 
Eq.(\ref{action_leading}) gives $B\sim 928.714$ while the thin-wall approximation Eq.(\ref{gbsoln}) gives $B_{tw}\sim 1019.82$. The relative difference is about $9.8\%$ due to the
finite difference of $V_+$ and $V_-$. (So the bubble wall is not ``thin" enough for this potential.)
\end{itemize}

As  $H$ is decreased from a large value, the tunneling type goes from case (IV) to case (II) and finally to case (I). The corresponding exponent $B$ for different cases are plotted in Figure \ref{power} as a function of $H$.
It is straightforward to see the dependence on the shape of potential, say, by varying $\lambda=\lambda_{\pm}$ and fixing $\Delta \phi_-=0.007$, $\Delta \phi_+=0.005$,
$V_T=8\times 10^{-7}$, we can also see the transition among the three cases. Notice the combination $x I(1/x)$ is monotonically increasing function of $x$, so the conditions can be solved easily in $\lambda$, 
\begin{itemize}
\item For $\lambda < 8.4257 \times 10^{-8}$ we have case (IV).
\item For $ 8.4257 \times 10^{-8}\leq \lambda < 1.23727 \times 10^{-7}$ we have case (II). 
\item for $\lambda \geq  1.23727 \times 10^{-7} $ we have case (I). 
\end{itemize}
Therefore case (IV), which is close to HM tunneling, happens for a ``flatter" barrier (small $\lambda$) while case (I), of which thin-wall tunneling is a special case, happens for a
``sharper" barrier (large $\lambda$).

We may write the factor $B$ (\ref{bounce1}) as composed of two terms,
\begin{equation}
B 
=[S_E(\phi_{f+}) - S_E(\phi_{+})] + [S_E(\phi_{bounce}) - S_E(\phi_{f+})]
\label{B2terms}
\q
where $\phi_T \ge \phi_{f+} \ge \phi_+$. 
For cases (I) and (II), we have $\phi_{f+}=\phi_+$, so the first term vanishes. 
For cases (III) and (IV), $\phi_{f+} > \phi_+$, so both terms contribute.
In the limit $\phi_{f+}  \rightarrow \phi_T$, $\phi(\xi) \rightarrow \phi_T$, so the second term vanishes
and the resulting formula reduces to the HM limit. The picture is consistent with the interpretation in \cite{Goncharov,Starobinsky,GLM}. However, as we have pointed out, corrections and back-reaction can introduce large corrections to the HM formula.
 
Brown and Weinberg \cite{Brown} showed that the CDL tunneling rate can be
derived by treating the field theory on a static patch of de Sitter space
as a thermal system.
In this thermal system, tunneling does not need to occur from the bottom
of the false potential well.  Instead, tunneling proceeds by a
combination of thermal excitation part way up the
barrier followed by quantum tunneling through the barrier.  The tunneling
rate is a thermal average of the energy-dependent quantum tunneling rates \cite{Affleck},
\begin{equation}
\Gamma \sim \int dE e^{-(E - E(\phi_+))/T} e^{-J(E)}
\label{BM_Tunneling_Rate}
\q
where $E$ is the energy of the field configuration.  Here the quantum
tunneling rate is given by WKB approximation $J(E) =2\int_{\phi_{f+}}^{\phi_{t-}}
d\phi \sqrt{2(V(\phi)-E)}$ where
the integral is along a Euclidean path from one classical turning point
$\phi_{f+}$ to the other classical turning point $\phi_{t-}$.  This integral is
dominated by the energy $E_*$ that maximizes the integrand.  
Using the saddle-point approximation it follows that
\m
\frac{1}{T} & = & 2 \int_{\phi_{f+}}^{\phi_{t-}} d\phi
\frac{1}{\sqrt{2(V(\phi)-E)}} \nonumber \\
 & = & 2 \int_{\phi_{f+}}^{\phi_{t-}} d\phi \frac{1}{\sqrt{(\frac{d \phi}{d
\tau})^2}} \\
 & = & 2 \Delta \tau
\n
since the integral is along a solution to the Euclidean equations of
motion. Fixing $T=T_H$ determines $E_*$ (and so $\phi_{f+}$ and $\phi_{t-}$). 
This yields $J(E_*) = S_E(\phi_{bounce}) - S_E(\phi_{f+})$.
This result also implies that the tunneling takes a Euclidean time $\Delta \tau = 1/ 2 T_H$ and  
\begin{equation}
S_E(\phi)=\int_{-\frac{1}{2T_H}}^{\frac{1}{2T_H}} E(\phi) d \tau \simeq E/T_H 
\q
That is, $E_*/T_H = S_E (\phi_{f+})$ and $E(\phi_+)/T_H = S_E (\phi_{+})$. 
Using this result in the saddle-point approximation to $(\ref{BM_Tunneling_Rate})$, one finds
\begin{equation}
\Gamma \sim e ^{-(S_E(\phi_{f+}) - S_E(\phi_{+}))} e^{-(S_E(\phi_{bounce}) - S_E(\phi_{f+}))}
\label{BBW}
\q
which reproduces the standard CDL tunneling rate.  
So our result (\ref{B2terms}) agrees with this result of Ref.\cite{Brown}: the first term in (\ref{B2terms}) corresponds to a thermal (i.e., Gibbons-Hawking temperature) fluctuation from the false vacuum (at $\phi_+$) part way up the barrier to $\phi_{f+}$ and the second term in (\ref{B2terms}) corresponds to a quantum tunneling.  Note that the derivation of Ref.\cite{Brown} assumes a fixed de Sitter background., while the back-reaction (which can be large) is included in our derivation. In this sense, our result is a big improvement. In Sec. 5.1, we calculated the $S_E(\phi_{bounce})$ and determined 
$\phi_{f+}$ for a given triangular potential. This allows us to compare quantitatively the contribution from the GH thermal effect to the contribution from the quantum effect in such a tunneling.   

\section{Thermal Tunneling}

At finite temperature, the transition from $V_+$ to $V_-$ can also occur via thermal fluctuations \cite{Linde}.  So there is another contribution due entirely to the GH temperature fluctuation. The dominant thermal tunneling process is the formation of $O(3)$ symmetric bubbles of $V_-$ that minimize the change in entropy $\Delta S$.  Let us evaluate the rate of this process and compare it to the above tunneling process.
The thermal tunneling rate is given by
\begin{equation}
e ^ {\Delta S} = e ^ {\Delta F / T}
\q
where $\Delta F$ is the change in the free energy.  We can think of the free energy as a three-dimensional action
\begin{equation}
\label{deltaF}
\Delta F = S_3 = \int d^3 x \bigg( \frac{1}{2}(\partial \phi) ^2 + V(\phi) \bigg)
\q
for a scalar field with a standard kinetic term.  In this case the equations of motion are
\m
\phi''+\frac{2 r'}{r} \phi'&=&\frac{dV}{d\phi} \label{eqphi3} \\
r'^2&=&1+\frac{ r^2}{3 M_p^2}  (\frac{1}{2} \phi'^2-V).
\label{S3_eqr}
\n

We restrict our attention to the symmetric triangle potential as before.  In the large $V_T$ limit considered above, the general solution to $(\ref{S3_eqr})$ is
\begin{equation}
\phi = \frac{\lambda}{H ^2} \big(A + \frac{H \xi}{2} \cot (H \xi) + B \cot (H \xi) \big)
\q
Like the Euclidean case discussed above, there are also four types of thermal tunneling solutions:
\begin{itemize}
\item Case I. $\phi(\xi)$ reaches both $\phi_+$ and $\phi_-$.
\item Case II. $\phi(\xi)$ reaches $\phi_+$ but not $\phi_-$.
\item Case III. $\phi(\xi)$ reaches $\phi_-$ but not $\phi_+$.
\item Case IV. $\phi(\xi)$ reaches neither $\phi_+$ nor $\phi_-$.  
\end{itemize} 
We will begin our discussion with case IV as it is the simplest.

\subsection{Case IV}
In this case, we need to solve $(\ref{eqphi3})$ with the boundary conditions $(\ref{boundary_condition_4})$.  The solution to these equations is
\m
\phi_1(\xi) &=& \phi_T+\frac{\lambda}{H ^2} \frac{H \xi}{2} \cot (H \xi) \\
\phi_2(\xi) &=& \phi_T-\frac{\lambda}{H ^2} \bigg(\frac{H \xi}{2} \cot (H \xi) - \frac{\pi}{2} \cot (H \xi)\bigg) \\
\xi_T &=& \frac{\pi}{2 H}
\label{S3_case_IV_solution}
\n
This solution is valid if
\begin{equation}
\frac{H ^ 2 \Delta \phi_+}{2 \lambda} < 1.
\label{S3_case_IV_condition}
\q
Comparing $(\ref{S3_case_IV_condition})$ to $(\ref{case_IV_condition})$ we see that the existence of a case (IV) Euclidean tunneling solution implies the existence of a case IV
thermal tunneling solution and that some potentials will admit a case IV thermal tunneling solution and a case (II) Euclidean tunneling solution.

Evaluating $(\ref{deltaF})$ using $(\ref{S3_case_IV_solution})$ we find that, for a specific background temperature,
\begin{equation}
S_3 / T_H = \frac{8 \pi \Delta V_+}{3 H^4} + O \bigg( \frac{\lambda ^2 M_p ^ 6}{V_T ^3} \bigg).
\label{S3_case_IV}
\q
To compare to the HM bounce $(\ref{BHM0})$ we find the ratio at leading order
\begin{equation}
\frac{B_{HM}}{S_3 / T_H} \sim 1.
\q
Based on the discussion in the last section, this result is expected and consistent with \cite{Weinberg:2007}. 
One should compare this thermal tunneling to the stochastic tunneling process of \cite{Goncharov,Starobinsky,GLM}.


\subsection{Case II}
Case III tunneling only occurs when $\lambda_+ \neq \lambda_-$ as in the Euclidean case. Since it does not show any 
special new feature, we shall skip this case.

In case II, we need to solve $(\ref{eqphi3})$ with the boundary conditions $(\ref{boundary_condition_2})$.  Imposing the condition $\phi_1 ' (0) = 0$ we find
\m
\phi_1(\xi) &=& \frac{\lambda}{H ^2} \bigg( A_1 + \frac{H \xi}{2} \cot (H \xi) \bigg) \\
\phi_2(\xi) &=& \frac{\lambda}{H ^2} \bigg(A_2 - \frac{H \xi}{2} \cot (H \xi) + B \cot (H \xi)\bigg) \\
\label{S3_case_II_solution}
\n
As in the Euclidean case we can eliminate $A_1$ and $A_2$ by rewriting the boundary conditions $(\ref{boundary_condition_2})$ as
\m
-\frac{H \xi_T}{2} \cot (H \xi_T) + \frac{H \xi_+}{2} \cot ^ 2 (H \xi_+) + B (\cot (H \xi_T) - \cot  (H \xi_+)) &=& \frac{H^2 \Delta \phi_+}{\lambda} \\
\cot (H \xi_T) - H \xi_T \csc ^ 2 (H \xi_T) - B \csc ^ 2 (H \xi_T) &=& 0 \\
\frac{1}{2} \cot ( H \xi_+) - \frac{H \xi_+}{2} \csc ^ 2 (H \xi_+) - B \csc ^ 2 ( H \xi_+) &=& 0
\n
which we can solve numerically for $B$, $\xi_T$, and $\xi_+$ as functions of $\frac{H^2 \Delta \phi_+}{\lambda}$.  We can then use these solutions in the original 
equations $(\ref{boundary_condition_2})$ to determine $A_1$ and $A_2$.

For our solution to be consistent, we need $\phi_1 (0) - \phi_T < \Delta \phi_-$ or equivalently
\begin{equation}
I_{S_3} \bigg(\frac{H ^ 2 \Delta \phi_+}{\lambda} \bigg) = \frac{1}{2} - \frac{H \xi_T}{2} \cot(H \xi_T) < \frac{H ^ 2 \Delta \phi_-}{\lambda}.
\label{S3_case_II_condition}
\q
Numerically we find that $I_{S_3}(\frac{H ^ 2 \Delta \phi_+}{\lambda}) > \frac{H ^ 2 \Delta \phi_+}{\lambda}$ so case II thermal tunneling is always from $V_+$ to some $V_F$ 
where $V_+ > V_F > V_-$ as we saw above in case (II) Euclidean tunneling.

\subsection{Case I}
When $(\ref{S3_case_IV_condition})$ and $(\ref{S3_case_II_condition})$ are both violated we have case I thermal tunneling.  Here we need to solve $(\ref{eqphi3})$ 
with the boundary conditions $(\ref{boundary_condition_1})$ numerically if the thin-wall approximation is not valid.  In general the solutions have the same qualitative features
as in case I Euclidean tunneling, but the Euclidean bounce $B$ is smaller than $S_3/T_H$.

Thin-wall CDL tunneling is a special case of I.  The thin-wall approximation greatly simplifies calculations so for the remainder of this section we consider a general 
potential that satisfies the thin-wall conditions.
The difference between the CDL and thermal tunneling rates has to do with $R^4$ space in CDL versus 
$R^3 \times S^1$ space in the thermal case, where $S^1$ corresponds to $\beta = 1/T= 2\pi/H$. 
In the thin-wall approximation we can divide the integration in \ref{deltaF} into three parts. For a bubble of radius $r$ we have $S_{3,out} = 0$ and 
$S_{3,wall} = 4 \pi r ^ 2 \tau$.
Using Eq.(\ref{echi}), the integral inside the bubble becomes
\begin{equation}
 S_{3,in}(\phi)=-{24\pi\over\kappa}\int_0^r d{\tilde r}(1-\kappa V(\phi) {\tilde
r}^2/3)^{1/2}. 
\q
The free energy is given by 
\begin{equation}
S_3 =  4  \pi r^2 \tau +\frac{12 \pi}{\kappa} 
\[r\(1- r^2/R_+^2\)^{1/2}  + R_+ \arcsin \bigg(\frac{r}{R_+}\bigg) - r\(1- r^2/R_-^2\)^{1/2}  - R_- \arcsin \bigg(\frac{r}{R_-}\bigg) \]. \label{s3}
\q

Maximizing $S_3$, we find $R$ given by
\begin{equation}
{1\over R^2}={1 \over (R_0^{(3)})^2}+{(H_+^2 + H_-^2)\over 2}+{\kappa^2\tau^2\over 36}
\q
where $R_0^{(3)}= 2 \tau/\epsilon$. 
In the $R \ll R_+$ case; this yields
\begin{equation} 
S_3 =  4  \pi R^2 \tau  - 4 \pi R^3 \epsilon/3
\q
For critical size radius $R = R_0^{(3)}$, we have
\begin{equation}
S_3/T= 32 \pi^2 \tau^3/3 \epsilon^2 H
\q
To see which path dominates, we compare this to $B=S_E$,
\begin{equation}
\frac{B}{(S_3/T)} = \frac{27}{64} \frac{ R_0^{(4)}}{R_+}
\q
Since $R \ll R_+$, we see that the Euclidean $S^4$ bounce $B$ is smaller so it dominates.

For large, $R \lesssim R_+$, 
\begin{equation}
S_3/T = 8 \pi^2 \tau/H^3
\q
so we see that $B$ (\ref{BLR}) is smaller than $S^3/T$ by a factor of four,
\begin{equation}
\frac{B}{(S_3/T)} = \frac{1}{4}
\q
It is easy to understand the origin of this factor of four.
In CDL, we have $S^3$ with size $2 \pi^2$ versus $S^2 \times S^1$ with size 
$4 \pi \times 2 \pi = 8 \pi^2$ in the thermal case. 
This just shows that the $O(4)$ symmetry lowers the Euclidean action $B$ and 
so yields the correct answer. 

Note that this is true only for $B \gg 1$.
Recall that
\begin{equation}
\Gamma = A_0 e^{-B} + A_T e^{-S_3/T} + . . . \sim A_1 e^{-B} + A_T e^{-4B}
\q
If this condition is not satisfied, say when $B \gtrsim 1$, then this factor of four difference may be overcome by the difference in the prefactors $A_0$ and $A_T$. 
Callan and Coleman \cite{Callan} show that, for bounce $B$,
\begin{equation}
 \Gamma =  \(\frac{B}{2\pi}\)^2 \(\frac{\det (-\partial^2 + V''(\phi_+)}{\det' (-\partial^2 + V''(\phi)}\)^{1/2} e^{-B}
\q
where the prime on the determinant implies that the four zero modes are removed, yielding the first factor. This prefactor is difficult to evaluate in general, so here we 
shall use dimensional arguments to find an order of magnitude estimate.

For $B \lesssim 1$, the above formula for CDL tunneling actually breaks down. Linde \cite{Linde} argued that
\begin{equation}
  \Gamma \sim T^4 \bigg(\frac{S_3}{2 \pi T}\bigg)^{3/2} e^{-S_3/T} 
\q
The difference is that here, the temperature $T$ is the Gibbons-Hawking temperature $T_H$, which strictly speaking is a gravitational, not thermal, effect.

This means that the potential $V$ should now include the finite temperature effect: $V(\phi_i) \rightarrow V(\phi_i, T_H)$.  The effective potential in de Sitter space has 
been calculated for some simple potentials in \cite{Allen}, \cite{Kane}.  As expected based on the GH temperature interpretation, the effective potential for scalar 
electrodynamics in de Sitter space calculated in \cite{Allen} shows the same behavior as one varies the inverse radius of de Sitter space as the effective potential in 
Minkowski space as one varies the temperature.  For a phenomenological potential (as in a 
slow-roll inflationary scenario), one may assume this finite temperature effect (and other quantum effects) is already built into the potential. However, in string theory 
applications to cosmology, where the effective potential can be calculated given a specific model, such finite GH temperature effects should be included. Here let us consider some specific examples.

\begin{figure}[t]
\centering
\includegraphics[width=10cm]{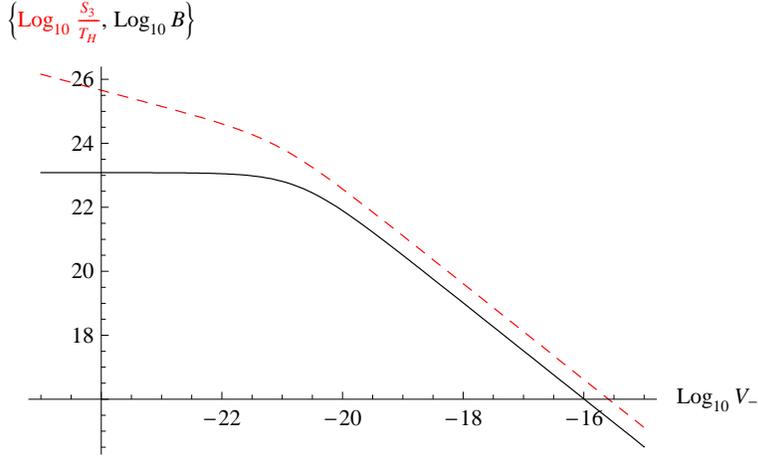}
\caption{Comparison of the Euclidean action and the free energy divided by the
GH temperature.  $\log_{10} B$ (the black solid curve) and $\log_{10} (S_3/T_H)$ 
(the red dashed curve) plotted against $\log_{10} V_-$ with $\tau = 10^{-10}$ and $\epsilon = 10^{-22}$ (in reduced Planck units) constant.  In the thin-wall approximation, CDL tunneling has an exponentially faster rate than thermal tunneling. We see that $B$ varies by 25 orders of magnitude. Tunneling is exponentially enhanced when the wavefunction is higher up (i.e., larger $V_T$) in the landscape.
}
\end{figure}



\section{Tunneling in the Cosmic Landscape}

We consider a type IIB compactification with the moduli stablized by a combination of fluxes and a
nonperturbative superpotential as in \cite{Kachru}.  These compactifications typically have a large number of axions $\phi_i$ corresponding to 
integrals of the four-form potential over each of the independent four-cycles.  Independent nonperturbative effects in the flux-induced superpotential give rise to a periodic
potential 
\begin{equation}
V(\phi_i) = M^4 e^{-S_{inst}^{i}} \bigg(1 - \cos \bigg(\frac{\phi_i}{f_i}\bigg)\bigg) +V(\rho)
\q  
where $f_i$ is the decay constant, $M$ is a natural mass scale (say, the string scale), and the $i$th instanton has action $S_{inst}^{i}$. $V(\rho)$ contains the potential coming from the moduli and 
includes $D$-term contributions.  The shift symmetries associated with these axions are broken by nonperturbative effects.  Including the effect of a single Euclidean D3-brane
for each independent four-cycle the axion potential becomes \cite{Easther}
\begin{equation}
V(\phi_i) = V_0(\phi_i) + \sum_i {\hat \alpha_i} \cos \bigg(\frac{\phi_i}{f_i}\bigg) + \sum_{ij} \beta_{ij} \cos \bigg(\frac{\phi_i}{f_i} - \frac{\phi_j}{f_j}\bigg) 
\q
where $V_0(\phi_i)$ is a smooth function of $\phi_i$ due to D-terms. Here, typical ${\hat \alpha_i}$ are exponentially small compared to the string scale.
It is easy to estimate $\tau_i$ along the $\phi_i$ direction.
Let $ \alpha_i = {\hat \alpha_i} + \sum_{j \ne i} \beta_{ij}$, so the height of the potential barrier in the $\phi_i$ 
direction is $2 \alpha_i$.
Then, using Eq.(\ref{tension}), we have
\begin{equation}
\tau_i = \int^{\phi_+}_{\phi_-} d \phi \sqrt{2 (2\alpha_i) \cos (\frac{\phi_i}{f_i})} = 
\frac{\pi}{2} \sqrt{\frac{\alpha_i}{2}} f_i
\q
and a crude estimate of the prefactor gives
\begin{equation}
\Gamma \simeq  \(\frac{B}{2\pi}\)^2 |V'' (\phi_t)|^2 e^{-B}
\q


where 
\begin{equation}
B = 2\pi^2\tau/H^3
\q
and $|V''(\phi_t)| = \alpha /f^2$, yielding
\begin{equation}
\Gamma = \frac{\pi^4}{8 H^9} \sum_{i=1}^N \frac{\alpha_i^3}{f_i^2} e^{-B_i}
\q

Tunneling can be fast if the axion decay constant $f$ is sufficiently small.  Cosmological upper bounds on the amount of axionic dark matter constrain $f$ to be below 
$10^{12}$ GeV, and astrophysical measurements of the cooling of red giants constrain $f$
to be at least $10^9$ GeV \cite{Svrcek}.  
For example if we take $f_i = 10^{10}$ GeV within this observationally preferred region and choose $\alpha_i = (2 * 10^{-7} M_p)^4$, then 
$\Gamma \sim 1$ for $H = 10^{-7} M_p$ and $N \sim 10$.  For smaller $f$ the tunneling rate $\Gamma$ can be order one for even smaller $H$.  
For model-independent axions in heterotic string theory $f$ is generically between $1.1 * 10 ^{16}$ GeV and the
reduced Planck mass \cite{Svrcek}.  If we choose $f_i \geq 1.1 * 10 ^{16}$ GeV for $N \leq 100$ there 
is no choice of $\alpha_i = \alpha$ and $H$ in the regime where effective field theory is valid ($\alpha \ll M_p^4$ and $H < M_p$) so that $\Gamma \sim 1$.

For relatively large $H$ or $T$, we have, instead
\begin{equation}
\Gamma \simeq  \(\frac{S_{i3}}{2\pi T}\)^2 |V'' (\phi_{it})|^2 e^{- S_{i3}/T}
\q
where $\phi_{it}$ is the value of $\phi_i$ at the top of the barrier.

	So it is reasonable to expect that the wavefunction of the universe tends to spread along some of the axionic directions. This what we expect for the QCD vacuum; that is,
	the wavefunction is a Bloch wave with angle $\theta_{QCD}$. As we go up in the cosmic landscape (say, turning on D-terms), tunneling will be faster and so the wavefunction
	will be Bloch wave-like in more of the axionic directions. If we start with the wavefeunction localized as a classically stable vacuum site, it will take time for the wavefunction to spread. This time will be shorter when we are higher up in the landscape and when there are more axionic directions present.

However, here $V(\phi_i)$ should be replaced by $V(\phi_i, T)$.
Instead of calculating this, we can make another estimate.
In quantum mechanics, tunneling through a barrier becomes less suppressed when the energy of the particle increases.
This happens when the particle is in a thermal bath with rising temperature. In quantum field theory, thermal effects typically lift the potential in a way such that 
the tunneling becomes faster, until the barrier disappears (i.e., the tunneling probability approaches unity). 

For $\lambda \phi^4$ theory, with 
$$V(\phi,T=0) = -\frac{m^2}{2} \phi^2 + \frac{\lambda}{4!}\phi^4 $$
we have, for high temperature $T$,
\begin{equation}
V(\phi,T_H) =\frac{1}{2}\bigg(\frac{\lambda T_H^2}{24} - {m^2}\bigg) \phi^2 +  \frac{\lambda}{4!}\phi^4 + ...
\q 
so the effective mass term is no longer tachyonic for $T > T_c$, where
the critical temperature is given by $T_c^2 \simeq 24 m^2/ \lambda$.

For any direction in the moduli space, if  
\begin{equation}
T_H^2 > \frac{24 m_i^2}{ \lambda_i},   \quad \quad i=1,2,.... , d
\q

Expanding the above potential about a maximum (top of a barrier),
we have $$V(\phi) = -\frac{\alpha}{2 f^2} \phi^2 + \frac{\alpha}{4! f^4}\phi^4 + ...$$
we see that, if in any direction,
\begin{equation}
24 f_i^2 < T_H^2
\q
then there is no barrier in that direction so the wavefunction is coherent along that direction.
Even if this condition is not satisfied, we see that the finite $T_H$ effect will enable to wavefunction to be coherent along some direction quickly, approaching a 
wavefunction similar to a Bloch wave. 

In fact, compactification and moduli stabilization usually introduces a term like $H^2\phi^2$, which will tend to remove a lot of barriers. This term arises from Gauss' Law.
 Actually, this is the term that causes the $\eta$ problem for slow-roll inflation. 
 
In a single dimension at low scales these effects are probably negligible without a large degree of fine-tuning.  For example consider a 
 de Sitter vacuum in flux compactification in string theory with all moduli fixed.  In the KKLT model  \cite{Kachru}, a highly warped type IIB compactification with nontrivial fluxes is stabilized using 
 non-perturbative effects from Euclidean D-branes and gaugino condensation.  The resulting anti-de Sitter minimum for the imaginary part of the volume modulus $\sigma$ is uplifted
  to a de Sitter minimum by adding a small number of $\overline{D3}$ branes.  The presence of the $\overline{D3}$ branes induces a term $D / \sigma ^3$ where $D$ depends on the warp 
  factor and the number of $\overline{D3}$ branes.  The potential is:
  \begin{equation}
  V = \frac{a A e ^{-a \sigma}}{2\sigma ^ 2} \bigg(\frac{1}{3} \sigma a A e^{-a \sigma} + W_0 + A e^{-a \sigma} \bigg) +
  \frac{D}{\sigma ^3}
  \q
    In this model if one tunes the parameters to have the de Sitter minimum be the cosmological constant observed today the tunneling rate is exponentially long.  
  
  The Gibbons-Hawking temperature effects are unimportant in the KKLT model because in the regime where the calculation is under control ($\sigma \gg 1$), the 
  de Sitter vacuum is necessarily exponentially suppressed compared to the Planck scale.
  Increasing $D$ both increases the false vacuum and decreases the barrier. 
  The de Sitter vacuum disappears classically if the D-term uplifting is too large.  Since 
  the true vacuum is always Minkowski in this model HM tunneling dominates if the de Sitter vacuum is large.  The HM bounce is 
  $B_{HM} = 24 \pi ^2 M_p^4 \Delta V_+/V_T^2$.  Since $M_p^4/V_T \gg 1$ in this model, and $\Delta V_+ / V_T \ll 1$ only when the barrier is about to disappear classically,
  the GH temperature corrections are small in this case.  Numerical calculations approximating the maximum with a quartic potential and computing the finite temperature 
  corrections using the GH temperature also show that the region of parameter space in which the de Sitter minimum is present classically but unstable quantum mechanically 
  is extremely small.

\section{Summary and Remarks}

In this paper, we discuss the relation between CDL thin-wall tunneling and HM tunneling. The picture is in agreement with the qualitative understanding one has already, but the details allow us to obtain a quantitative understanding of the validity of each approximation and the sizes of the corrections. 
This result does allow us to say something about the stringy cosmic landscape. Here are a few comments that may be  relevant, in no particular order:

$\bullet$ When we are high up in the cosmic landscape, gravitational effects exponentially enhance the tunneling rate. Qualitatively, one may interpret this as a GH temperature effect, but the actually enhancement is exponentially larger than a naive estimate based on the usual temperature effect, due to the enhanced $O(4)$ symmetry of the Euclidean action (versus the $SO(3) \times U(1)$ symmetry in the finite temperature case).

$\bullet$ Above the GUT scale, quantum gravity effects become important and we have nothing to say in this case. Around or slightly below the GUT scale, the tunneling can easily be enhanced by so much that there is simply no exponential suppression at all and the semi-classical formula simply breaks down. This is in the single field case, with only one tunneling direction. In the multi-field ($d>1$) case, as we expect to be the situation in the cosmic landscape, there are typically many tunneling directions for any false vacuum, so one expects fast tunneling to take place.  

$\bullet$ As is well known (see Appendix A in Ref. \cite{Huang:}), for a generic spherical potential in $d>3$ spatial dimensions, the extra dimensions acts as an angular momentum like repulsive potential (with angular momentum $l = (d-3)/2$). So it is harder to form a bound state in higher dimensions, as is the case in the cosmic landscape. Instead of having a bound wavefunction that has to tunnel out, we may have a resonance like situation where the decay time can be very short. Even if the wavefunction is trapped, it is more likely to be weakly trapped, so the wavefunction has a long tail outside the classically allowed region, rendering less suppressed tunneling. 
 
$\bullet$ It is again well known that the finite temperature effect on the effective potential tends to decrease the barrier, allowing faster tunneling. In fact, compactification
and moduli stabilization usually introduces a term like $H^2\phi^2$, which will tend to remove a lot of barriers. This term, a finite volume effect, arises from the equivalent of
the Gauss's Law. (Actually, this is the term that causes the $\eta$ problem for slow-roll inflation \cite{Kachru:2003sx}) Here, a similar term arises due to the finite temperature
effect on the potential. One may re-interpret this effect as a finite volume effect, due to the presence of the Gibbons-Hawking horizon $1/H$.

$\bullet$ Even if the barriers are still present (that is, not completely lifted), we see that the finite $T_H$ effect will enable the wavefunction to be coherent along some
direction more quickly. This is particularly likely along the axionic directions since the effective potential is periodic or close to being periodic and the heights of the
barriers can be very low. In this situation, the wavefunction approaches a Bloch wave. Presumably, the QCD $\theta$ vacuum we live in today is described by such a Bloch wave.

$\bullet$ Since there are many vacua in the stringy landscape, resonance tunneling effects should play an important role in the landscape \cite{Tye:2006tg}. This also enhances the
tunneling among the vacua in the landscape. Such enhancement can be substantial. In particular, its role in the development of the Bloch wave is well known.

$\bullet$ With all these observations, one suspects that the wavefunction of the universe in the cosmic landscape
may be quite mobile; that is, it does not stay at any particular vacuum site long enough to allow eternal inflation. However, without a detailed knowledge of the structure of the
landscape (vacuum sites and their nearby neighbors as well as the barriers between them), it is difficult to make definitive statements. As we just pointed out, the wavefunction
most likely is spread out along the periodic directions (i.e., the axionic directions) as Bloch waves; so what happens to the aperiodic directions? If we treat such directions
randomly, i.e., as a random potential, one may borrow the insight obtained in condensed matter physics to argue that the wavefunction should be fully mobile along these directions as well \cite{Tye:2007ja}. This implies that eternal inflation in the landscape is very unlikely. 

This argument is based on the large dimensionality of the landscape and does not use the gravitational effects discussed in this paper; so it should apply all the way to very low vacuum energy density sites.

\vspace{0.7cm}

\noindent {\bf Acknowledgments}

\vspace{0.4cm}

We thank Adam Brown, Qing-Guo Huang, Gary Shiu and Erick Weinberg for valuable discussions.
This work is supported by the National Science Foundation under grant PHY-0355005.

\end{document}